\documentclass[5p,times]{elsarticle}

\usepackage[svgnames,dvipsnames,x11names, table]{xcolor}

\usepackage{acronym}
\usepackage{algorithm,algpseudocode}
\usepackage{amsmath}
\usepackage{appendix}
\usepackage{mathtools}
\usepackage{amsfonts}
\usepackage{amssymb}
\usepackage{bm}
\usepackage[font=small,labelfont=bf]{caption}

\usepackage{commath}
\usepackage{enumitem} 
\usepackage[c]{esvect}
\usepackage{float}
\usepackage{graphicx}
\graphicspath{{../}{./}}

\usepackage{hyperref}
\hypersetup{pdfauthor=author}

\usepackage{listings}
\usepackage{siunitx}
\usepackage{subcaption}
\usepackage{xspace}

\usepackage[normalem]{ulem}

\usepackage[capitalise]{cleveref}

\usepackage{color}

\usepackage{tikz}
\usetikzlibrary{shapes,arrows}
\usetikzlibrary{shapes.multipart}
\usepackage{adjustbox}

\usepackage{overpic}

\definecolor{mygray}{rgb}{0.5,0.5,0.5}
\definecolor{lightgray}{rgb}{0.95,0.95,0.95}

\usepackage{fancyvrb} 
\VerbatimFootnotes 

\definecolor{XcodeComments}{RGB}{00,116,00}
\definecolor{XcodeKeywords}{RGB}{170,13,145}
\definecolor{XcodeStringstyle}{RGB}{21,26,255}
\lstdefinestyle{pythonCode}{
  language=Python, 
  morekeywords={size_t},
  commentstyle=\color{XcodeComments}\fontseries{b}\fontshape{sl}\selectfont,
  keywordstyle=\color{XcodeKeywords}\fontseries{b}\selectfont,
  stringstyle=\color{XcodeStringstyle}\bfseries,
  emphstyle=\color{black}\bfseries
}

\journal{ArXiv.org}

\newcommand\mir{\textsc{Mirheo}\xspace}
\newcommand\aphros{\textsc{Aphros}\xspace}
\newcommand\lammps{\textsc{LAMMPS}\xspace}

\newcommand\pizd{\textit{Piz Daint}\xspace}

\newacro{CSCS}{Swiss National Supercomputing Centre}
\newcommand{\CSCS}{\ac{CSCS}\xspace}

\newacro{HPC}{High-Performance Computing}
\newcommand{\HPC}{\ac{HPC}\xspace}

\newacro{UQ}{uncertainty quantification}
\newcommand{\UQ}{\ac{UQ}\xspace}

\newacro{DPD}{dissipative particle dynamics}

\newacro{RBC}{red blood cell}
\acrodefplural{RBCs}{red blood cells}
\newcommand{\RBC}{\ac{RBC}\xspace}

\newacro{MS}{membrane skeleton}

\newacro{LWM}{Lim-Wortis-Mukhopadhyay}

\newacro{FSI}{flow-structure Interactions}

\newacro{SDE}{stomatocyte-discocyte-echinocyte}

\newacro{GPU}{graphics processing unit}
\acrodefplural{GPUs}{graphics processing units}
\newcommand{\GPU}{\ac{GPU}\xspace}

\newacro{MCMC}{Markov chain Monte Carlo}

\newacro{HMC}{Hamiltonian Monte Carlo}
\newcommand{\HMC}{\ac{HMC}\xspace}

\newacro{DRAM}{Delayed Rejection Adaptive Monte Carlo}

\newacro{TMCMC}{Transitional Markov \- Chain Monte Carlo}
\newcommand{\TMCMC}{\ac{TMCMC}\xspace}

\newacro{NS}{nested sampling}
\newcommand{\NS}{\ac{NS}\xspace}

\newacro{CMA-ES}{Co\-vari\-ance Matrix Adaption Evolution Strategy}
\newcommand{\CMAES}{\ac{CMA-ES}\xspace}

\newacro{ES}{evolution strategy}
\newcommand{\ES}{\ac{ES}\xspace}

\newacro{LMCMA}{Limited Memory Covariance Matrix Adaption}

\newacro{DEA}{Differential Evolution Algorithm}

\newacro{Rprop}{Resilient Backpropagation}

\newacro{BASIS}{Bayesian Annealed Sequential Importance Sampling}
\newcommand{\BASIS}{\ac{BASIS}\xspace}

\newacro{RL}{Re\-in\-force\-ment Learning}
\newcommand{\RL}{\ac{RL}\xspace}

\newacro{DLD}{deterministic lateral displacement}

\newacro{MAP}{maximum a posteriori}

\newacro{CG}{coarse-grained}
\newcommand{\CG}{\ac{CG}\xspace}

\newcommand{\PM}{\bm{\vartheta}}    

\newcommand{\X}{\bm{x}}
\newcommand{\vX}{\vv{\X}}

\newcommand{\Y}{y}
\newcommand{\vY}{\bm{\Y}}

\newcolumntype{M}[1]{>{\centering\arraybackslash}m{#1}}
\newcolumntype{N}{@{}m{0pt}@{}}
\newcommand{\GIVEN}{\hspace{0.1em}   |  \hspace{0.1em}}

\definecolor{light-gray}{gray}{0.55}

\begin{document}

\begin{frontmatter}

\title{Korali: Efficient and Scalable Software Framework for\\ Bayesian Uncertainty Quantification and Stochastic Optimization}

\author[cselab]{Sergio M. Martin}
\author[cselab]{Daniel W{\"a}lchli}
\author[cselab]{Georgios Arampatzis}
\author[cselab]{Athena E. Economides}
\author[cselab]{Petr Karnakov}
\author[cselab,harvard]{Petros Koumoutsakos\corref{corAuth}}

\cortext[corAuth]{Corresponding author}\ead{petros@ethz.edu}

\newcommand{\cselab}{Computational Science and Engineering Laboratory, ETH Z\"{u}rich, CH-8092, Switzerland.}

\newcommand{\harvard}{Institute of Applied Computational Sciences, Harvard University, Cambridge, MA, USA.}

\address[cselab]{\cselab}
\address[harvard]{\harvard}

\begin{abstract}

We present Korali, an open-source framework for large-scale Bayesian uncertainty quantification and stochastic optimization. The framework relies on non-intrusive sampling of complex multiphysics models and enables their exploitation for optimization and decision-making. In addition, its distributed sampling engine makes efficient use of massively-parallel architectures while introducing novel fault tolerance and load balancing mechanisms. We demonstrate these features by interfacing Korali with existing high-performance software such as \aphros, \lammps (CPU-based), and \mir (GPU-based) and show efficient scaling for up to 512 nodes of the CSCS Piz Daint supercomputer. Finally, we present benchmarks demonstrating that Korali outperforms related state-of-the-art software frameworks. 
 
\end{abstract}


\end{frontmatter}

\section{Introduction}

Over the last thirty years, \HPC architectures have enabled high-resolution simulations of physical systems ranging from atoms to galaxies. \HPC has also reduced the cost and turnaround time of such simulations, making them invaluable predictive and design tools across all fields of science and engineering. Multiple simulations at resolutions that would have been impossible a decade ago are routinely employed in optimization and design. The  transformation of these simulations into actionable decisions requires the quantification of their uncertainties. In recent years, \HPC has become central in the way that we conduct science with massive amounts of data. Such data are used to develop and calibrate physical models as well as to quantify the uncertainties of their predictions. The integration of data and physical models has a history of over 300 years, dating back to Laplace and Copernicus and to the framework known as Bayesian inference. However, due to its computational cost, the application of Bayesian inference has been, until recently, limited to simple models or through inexpensive approximations. 

Bayesian inference requires sampling of distributions with dimensionality greater or equal to the number of model parameters. The sampling necessitates numerous evaluations, making the process computationally demanding, particularly when the underlying model requires hundreds of compute hours per evaluation. Moreover, special care is necessary to develop sampling algorithms that harness the capabilities of modern supercomputers~\cite{ashby:2010}.
The sampling involved in Bayesian inference serves as a bridge to 
stochastic optimization algorithms \cite{Hansen2003} that aim to identify the probability distribution of the parameters that maximize a particular cost function. distribution. Stochastic optimization algorithms such as CMA-ES, MBOA and Natural Gradient Optimization \cite{Kern2004,Akimoto:2014} are non-intrusive and rely on handling the physical models through input/output relations.

The need for efficient deployment of optimization and uncertainty quantification algorithms has motivated to the development of several statistical frameworks \cite{uqlab:page,  easyvvuq:page, abcsysbio2010, aptmcmc, psuade:page, queso2011, stan:page}. However, to the best of our knowledge, only a few such frameworks are well-suited for deployment in massively parallel computer architectures \cite{Hadjidoukas2015, dalbey2020dakota}. In this paper, we present Korali, a new framework for Bayesian \UQ and optimization. The framework enables efficient large-scale sampling while providing mechanisms for fault-tolerance, load balancing, and reproducibility, which are essential requirements for emerging supercomputers \cite{dongarra:Exascale}. We demonstrate these capabilities guided by three motivating studies. First, a high-dimensional optimization study of the shape of fluid transporting pipes.  Second, a hierarchical Bayesian analysis of the dissipation parameter of human \RBC membranes. Lastly, a Bayesian study on the parameters of a \CG water model. In addition, we provide a synthetic benchmark that compares the performance of Korali with that of other state-of-the-art frameworks for optimization on up to 512 nodes of the CSCS Piz Daint supercomputer.

The rest of this paper is organized as follows: in \cref{sec:unified}, we present the principles behind the unifying framework; in \cref{sec:design}, we present the framework's design; in \cref{sec:experiments}, we present the results of three experimental cases; in \cref{section:related}, we discuss state of the art \UQ frameworks and compare their efficiency, and; in \cref{sec:conclusion}, we present conclusions and future work.

\section{Unified Approach to Optimization and Sampling}\label{sec:unified}

We designed Korali to exploit the common patterns in optimization and sampling while exposing a unifying interface. Consider a computational model, represented by a function $f$, that depends on parameters $\PM$ with unknown values and possibly other parameters $\X$ with known values. We wish to infer the values for the parameters $\PM$ such that the model evaluated at known parameters $\X_i$ will approximate given values $\Y_i$ for $i=1,\ldots,N$. The variables $\Y_i$ are called the \textit{data} and usually correspond to experimental measurements. Since the measurements are typically affected by random errors, we introduce a probabilistic model that associates the measurements $\Y_i$ with the model evaluations $f(\PM,\X_i)$. This model is represented by a probability density function $p(\vY \GIVEN \PM; \vX)$ where $\vY=\{\Y_1,\ldots,\Y_N\}$ and $\vX=\{\X_1,\ldots,\X_N\}$. For fixed $\vX$ and $\vY$ the density is a function of the parameters $\PM$ and is called the \textit{likelihood} function. Furthermore, any information on the parameters $\PM$ that is known prior to observing any data is encoded in a density function $p(\PM)$. The distribution of the parameters $\PM$ conditioned on the known values $\Y_i$ is given by Bayes' theorem: $p(\PM \GIVEN \vY;\vX) \propto p(\vY \GIVEN \PM; \vX) p(\PM)$. The posterior density function can either be optimized or sampled.

By optimizing the posterior density we obtain a single value for the vector $\PM$ that represents the value of the parameters with the highest probability. If the derivatives of the posterior with respect to $\PM$ are available, a local optimization method can be used, e.g., the Adam algorithm \cite{Adam2015}. Otherwise, a derivative free optimization algorithm can be used, e.g., \ES algorithms. At every iteration, this type of algorithms draw samples from a parametrized distribution family and rank them from highest to lowest value. For example, the CMA-ES \cite{Hansen2003} uses this ranking and the history of the evolution to update the mean and the covariance matrix of a normal distribution. In the limit, the normal distribution converges to a Dirac distribution located at an optimal value. The covariance at each iteration can be interpreted as scaled approximation of the Hessian of the objective function at the mean of the normal distribution.

If we sample the posterior distribution instead of optimizing it, we obtain a set of samples that represent the volume of the distribution in the parameter space. This extended information, compared to the single parameter obtained with optimization, allows us to estimate the uncertainty in the inference of the parameters. Moreover, the uncertainty can be propagated to quantities of interest of the computational model, assessing this way the uncertainty in the predictions of the model. If derivatives are available, algorithms like \HMC \cite{Gelman2014} can be utilized that accelerate the sampling and are efficient in high-dimensional spaces. In the opposite case, derivative-free algorithms, similar to Metropolis-Hastings, \NS \cite{feroz2009} or \TMCMC \cite{Wu2018} can be utilized. In particular, the \NS and the \TMCMC algorithms have the benefit of parallel evaluation of samples, allowing the acceleration of sampling of problems that involve computationally demanding models. 

The common pattern between optimization and sampling algorithms is the iterating cycle of evaluation of the posterior density function. Additionally, for algorithms like \ES, \NS and \TMCMC many function evaluations can be executed in parallel within the same iteration.

\section{Framework Design}\label{sec:design}

The framework is specially tailored for the execution of massively parallel population-based algorithms that rely on generating and evaluating sets of parameters (samples) iteratively. At each iteration, a set of $n$ samples $S_i$ are evaluated by a statistical model $\ell$. The statistical model to use is given by the choice of \textit{problem}, \textit{e.g.}, \textit{Optimization}, to search the optimum of an objective function; \textit{Sampling}, to sample an unnormalized density function, and; \textit{Bayesian Inference}, to sample the posterior distribution and uncertainty of an inverse Bayesian problem. The statistical model evaluation may require the execution of a computational model ($f$) representing, \textit{e.g.,} the simulation of a physical system.

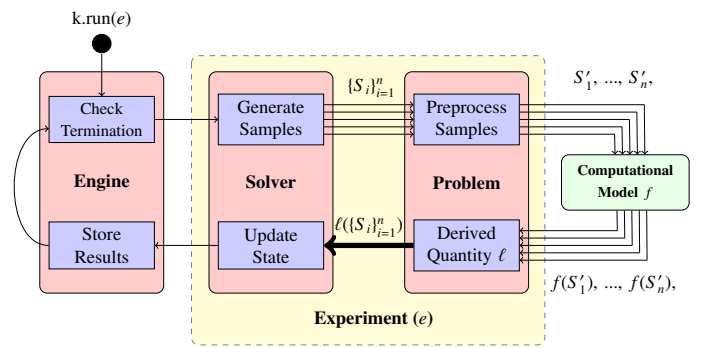
\begin{figure}[htb]
\begin{center}
    \begin{adjustbox}{width=\columnwidth}
    \hspace{-40pt} 
\pgfdeclarelayer{background}
\pgfdeclarelayer{foreground}
\pgfsetlayers{background,main,foreground}

\tikzstyle{subModule}=[draw, fill=blue!20, text width=5em, text centered, minimum height=2.5em]
\tikzstyle{module} = [subModule, text width=6em, fill=red!20, minimum height=12em, rounded corners]
\tikzstyle{worker} = [draw, fill=orange!50, text width=5em, text centered, minimum height=3.0em]
\tikzstyle{rank} = [draw, fill=orange!50, text width=5em, text centered, minimum height=1.5em]
\tikzstyle{model} = [draw, fill=green!10, text width=6.2em, text centered, minimum height=3em, rounded corners]
\def\blockdist{2.3}
\def\edgedist{2.5}

\tikzstyle{doc}=[%
draw,
thick,
align=center,
color=black,
shape=document,
minimum width=7mm,
minimum height=13.2mm,
shape=document,
inner sep=2ex,
]

\begin{tikzpicture}
    
    \node (problem) [module] {\textbf{Problem}};
    \path (problem.center)+(0,1.2) node (preprocess) [subModule] {Preprocess Samples};
    \path (problem.center)+(0,-1.2) node (postprocess) [subModule] {Derived Quantity $\ell$};

    \path (problem)+(-3.7,0) node (solver) [module] {\textbf{Solver}};
    \path (solver.center)+(0,1.2) node (sampleGen) [subModule] {Generate Samples};
    \path (solver.center)+(0,-1.2) node (updatedState) [subModule] {Update State};
    
    \path (solver)+(-3.2,0) node (engine) [module] {\textbf{Engine}};
    \path (engine.center)+(0,1.2) node (checkTermination) [subModule] {\small{Check Termination}};
    \path (engine.center)+(0,-1.2) node (resultsStorage) [subModule] {Store Results};
    \path [draw, ->] (resultsStorage.west) to [out=180, in=180] node [align=center] {} ([yshift=-5] checkTermination.west);
    
    \path (problem)+(+3,0) node (model) [model] {\footnotesize{\textbf{Computational Model}} $f$};
    
    \path [draw, ->] (checkTermination.north)+(0.0, +0.8) -- node (endDot.east) [align=center, yshift=30pt] {k.run($e$)} (checkTermination.north);
    \filldraw ( checkTermination.north)+(0.0,1.0) circle (5pt);
    
    \path (problem.south)+(-1.8,-0.5) node (ExpSubtitle) {\textbf{ Experiment ($e$)}};
    
    \begin{pgfonlayer}{background}
        \path (sampleGen.west |- problem.north)+(-0.5,0.3) node (a) {};
        \path (ExpSubtitle.south -| problem.east)+(+0.3,-0.2) node (b) {};
        \path[fill=yellow!20,rounded corners, draw=black!50, dashed] (a) rectangle (b);
    \end{pgfonlayer}

    \path (sampleGen.east)+(0.9, 0.6) node (preProcessText) {$\{S_i\}_{i=1}^n$};
    \path (preprocess.east)+(1.8, 0.75) node (preProcessText) {$S'_{1}, \, ..., \, S'_{n}, \, $};
    \path (postprocess.east)+(1.8, -0.75) node (preProcessText) {$f(S'_{1}), \, ..., \, f(S'_{n}), \, $};
    \path [draw, ->, line width=1.00mm] (postprocess.west) -- node [above] (preProcessText) {$\ell(\{S_i\}_{i=1}^n)$} (updatedState.east);  
        
    \path [draw, ->] (updatedState.west) -- node [xshift=-5, above, align=center] {} (resultsStorage.east);
    \path [draw, ->] (checkTermination.east) -- node [above, align=center] {} (sampleGen.west);    
    
    \pgfmathsetmacro{\n}{5}
    \foreach \i in {1,...,\n}
    {
      \path [draw, ->] ([yshift=-12 + \i*4.0pt] sampleGen.east) -- ([yshift=-12 + \i*4.0pt] preprocess.west);
      
      \path (preprocess.east)+(43.5 + \i*4.0pt, -12 + \i*4.0pt) node (collectPreNodeA\i) {};
      \path [draw, -] ([yshift=-12 + \i*4.0pt] preprocess.east) -- (collectPreNodeA\i.east);
      \path [draw, ->] (collectPreNodeA\i.east) -- ([xshift=-9.5 + \i*4.0pt] model.north) ;
      
      \path (postprocess.east)+(69 - \i*4.0pt, -12 + \i*4.0pt) node (collectPreNodeB\i) {};
      \path [draw, <-] ([yshift=-12 + \i*4.0pt] postprocess.east) -- (collectPreNodeB\i.east);
      \path [draw, -] (collectPreNodeB\i.east) -- ([xshift=15.5 - \i*4.0pt] model.south);
    }

\end{tikzpicture}
    \end{adjustbox}
\end{center}
\vspace{-10pt}
\caption{Generation-based workflow of the sampling engine. The \textit{solver} module generates \textit{samples} $S_i$. The \textit{problem} module pre-processes the samples into $S_i'$ and passes them to a computational model. The results of the computational model $f(S_i')$ are processed into $\ell(S_i)$ and the state of the solver is updated.}
\label{fig:engine_simple}
\end{figure}

To solve a given problem, the framework runs an \textit{experiment} (see \cref{fig:engine_simple}). Experiments are user-defined (for a detailed description of the user interface, see \ref{appendix:interface}) and contains the required configuration for the problem, the solver algorithm, and the \textit{variable space}. The variable space represents the range of values within which the solution is to be identified. Variables are uniquely identified by their name and can be restricted either through an upper and a lower bound, or described by a prior distribution.

Experiments run under the control of the framework's sampling engine. The engine will coordinate the exchange of samples between the experiment and the computational model until the solver terminates. \Cref{fig:engine_simple} shows the workflow of the engine when executing a given experiment. A \textit{generation} represents a cycle in which the experiment produces and evaluates a population of samples. We define a \textit{sample} $S_i$ as a particular selection of values within the feasible variable space.

\label{sec:unifying}

The first generation starts when the user runs \texttt{k.run(e)}, where \texttt{e} is the user-defined experiment object and \texttt{k} represents an instance of the engine. The first step in every generation is to check whether any of the defined termination criteria has been met. If not, the engine yields execution to the solver algorithm, which generates an initial population of samples $\{S_i\}_{i=1}^n$ and relays them to the problem module for pre-processing. During this stage, the samples are transformed to the correct input format for the computational model. The samples ($S_i'$) are then passed on to the computational model for evaluation ($\{f(S'_{i})\}_{i=1}^n$). 

Upon receiving the results from the computational model, the problem module calculates a derived quantity, e.g., the log-likelihood $\ell(S)$ for a problem of type Bayesian inference, and passes it back to the solver module. The solver uses these quantities to update its internal state and produce partial results, which serve as the basis for creating the next sample population during the next generation. 

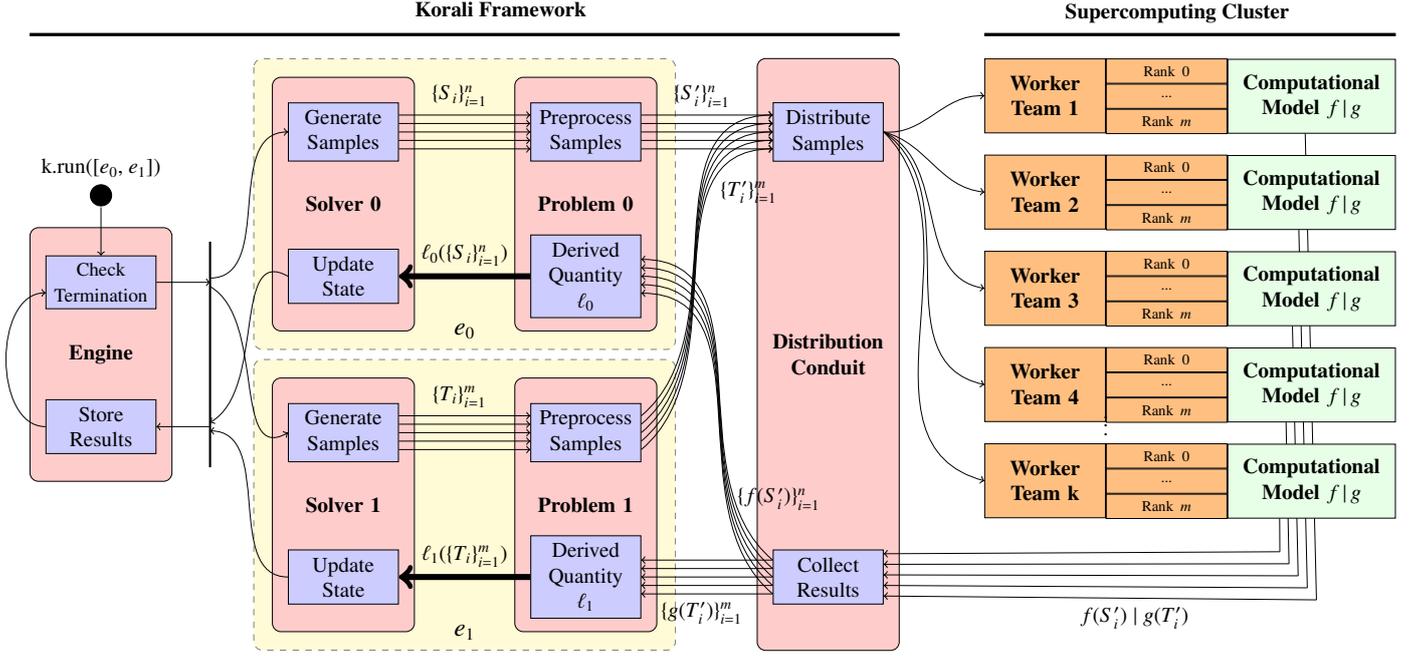
\begin{figure*}[htb]
\begin{center}
\begin{adjustbox}{width=\textwidth}
 \hspace{-40pt} 
\pgfdeclarelayer{background}
\pgfdeclarelayer{foreground}
\pgfsetlayers{background,main,foreground}

\tikzstyle{subModule}=[draw, fill=blue!20, text width=4.5em, text centered, minimum height=2.5em]
\tikzstyle{module} = [subModule, text width=6em, fill=red!20, minimum height=12em, rounded corners]
\tikzstyle{worker} = [draw, fill=orange!50, text width=5em, text centered, minimum height=3.5em]
\tikzstyle{rank} = [draw, fill=orange!50, text width=5em, text centered, minimum height=1.15em]
\tikzstyle{model} = [draw, fill=green!10, text width=7.2em, text centered, minimum height=3.5em]
\def\blockdist{2.3}
\def\edgedist{2.5}

\tikzstyle{doc}=[%
draw,
thick,
align=center,
color=black,
shape=document,
minimum width=7mm,
minimum height=13.2mm,
shape=document,
inner sep=2ex,
]

\begin{tikzpicture}
    
    \node (engine) [module] {\textbf{Engine}};
    \path (engine.center)+(0,1.2) node (checkTermination) [subModule] {\small{Check Termination}};
    \path (engine.center)+(0,-1.2) node (resultsStorage) [subModule] {Store Results};
    \path [draw, ->] (resultsStorage.west) to [out=180, in=180] node [align=center] {} ([yshift=-5] checkTermination.west);
    \path [draw, ->] (checkTermination.north)+(0.0, +0.8) -- node (endDot.east) [align=center, yshift=30pt] {k.run([$e_0$, $e_1$])} (checkTermination.north);
    \filldraw ( checkTermination.north)+(0.0,1.0) circle (5pt);

    
    \path (engine.center)+(1.8, 2) node (lineUp) {};
    \path (engine.center)+(1.8,-2) node (lineDown) {};
    \draw[line width=0.35mm] (lineUp) -- (lineDown);
    \path [draw, ->] (lineUp |- resultsStorage) -- (resultsStorage.east);
    \path [draw, ->] (checkTermination.east) -- (lineUp |- checkTermination);
    
    
    \path (engine)+(4,2.5) node (solver0) [module] {\textbf{Solver 0}};
    \path (solver0.center)+(0,1.2) node (sampleGen0) [subModule] {Generate Samples};
    \path (solver0.center)+(0,-1.2) node (updatedState0) [subModule] {Update State};
    
    \path (solver0.center)+(4.0,0)  node (problem0) [module] {\textbf{Problem 0}};
    \path (problem0.center)+(0,1.2)  node (preprocess0) [subModule] {Preprocess Samples};
    \path (problem0.center)+(0,-1.2) node (postprocess0) [subModule] {Derived Quantity $\ell_0$};
    
    \path (problem0.south)+(-2.0,0.0) node (ExpSubtitle0) {\large$e_0$};
    
    \begin{pgfonlayer}{background}
        \path (solver0.north west)+(-0.3,0.3) node (a0) {};
        \path (problem0.south east)+(+0.3,-0.3) node (b0) {};
        \path[fill=yellow!20,rounded corners, draw=black!50, dashed] (a0) rectangle (b0);
    \end{pgfonlayer}
    
    \path [draw, ->] ([yshift=2] lineUp |- checkTermination) to [out=0, in=180] (sampleGen0.west);
    \path [draw, ->] (updatedState0.west) to [out=150, in=30] ([yshift=2] lineUp |- resultsStorage);
    
    
    \path (engine)+(4,-2.5) node (solver1) [module] {\textbf{Solver 1}};
    \path (solver1.center)+(0,1.2) node (sampleGen1) [subModule] {Generate Samples};
    \path (solver1.center)+(0,-1.2) node (updatedState1) [subModule] {Update State};
    
    \path (solver1)+(4.0, 0) node (problem1) [module] {\textbf{Problem 1}};
    \path (problem1.center)+(0,1.2) node (preprocess1) [subModule] {Preprocess Samples};
    \path (problem1.center)+(0,-1.2) node (postprocess1) [subModule] {Derived Quantity $\ell_1$};
    
    \path (problem1.south)+(-2,0.0) node (ExpSubtitle1) {\large$e_1$};
    
    \begin{pgfonlayer}{background}
        \path (solver1.north west)+(-0.3,0.3) node (a0) {};
        \path (problem1.south east)+(+0.3,-0.3) node (b0) {};
        \path[fill=yellow!20,rounded corners, draw=black!50, dashed] (a0) rectangle (b0);
    \end{pgfonlayer}
    
    \path [draw, ->] ([yshift=-2] lineUp |- checkTermination) to [out=-30, in=210] (sampleGen1.west);
    \path [draw, ->] (updatedState1.west) to [out=180, in=0] ([yshift=-2] lineUp |- resultsStorage);
    
    
    \path (engine)+(12,0) node (conduit) [module, minimum height=28em]{\textbf{Distribution Conduit}};
    \path (conduit.center)+(0,3.7) node (distributeSamples) [subModule] {Distribute Samples};
    \path (conduit.center)+(0,-3.7) node (collectResults) [subModule] {Collect Results};
    
    \pgfmathsetmacro{\n}{5}
    \foreach \i in {1,...,\n}
    {
      \path [draw, ->] ([yshift=-12 + \i*4.0pt]sampleGen0.east) -- ([yshift=-12 + \i*4.0pt]preprocess0.west);
      \path [draw, ->] ([yshift=-12 + \i*4.0pt]preprocess0.east) -- ([yshift=-12 + \i*4.0pt]distributeSamples.west);
      
      \path [draw, ->] ([yshift=-12 + \i*4.0pt]sampleGen1.east) -- ([yshift=-12 + \i*4.0pt]preprocess1.west);
      \path [draw, ->] ([yshift=-12 + \i*4.0pt]preprocess1.east) to [out=40,in=180] ([yshift=-12 + \i*4.0pt]distributeSamples.west);
      
      \path (conduit.east)+(2.4, +5.9 - \i*1.6) node (worker\i) [worker] {\textbf{Worker Team \ifthenelse{\i=5}{k}{\i}}};
      
      \ifthenelse{\i=5}{\path (worker\i.north east)+(0,0.4) node (manyWorkers) {$\vdots$};}{}
      
      \path (worker\i.north east) node[anchor=north west] (rank_\i_0) [rank] {\scriptsize{Rank 0}};
      \path (rank_\i_0.south) node[anchor=north] (rank_\i_1) [rank] {\scriptsize{...}};
      \path (rank_\i_1.south) node[anchor=north] (rank_\i_m) [rank] {\scriptsize{Rank $m$}};
      
      \path [draw, ->] (distributeSamples.east) to [out=-10*\i + 5, in=180 - 3*\i] node [xshift=10 +  \i*1, yshift=32 - \i*16] (sampleSetTitle\i) {} (worker\i.west) ;
      
      \path (rank_\i_1.east) node[anchor=west] (model\i) [model] {\textbf{Computational Model} $f \, | \, g$ };
      \path (collectResults.center)+(230 -\i*4.3pt,-0.58 + 0.2*\i) node (collectPreNodeA\i) {};
      \path [draw, -] (model\i.south)+(-\i*3pt,0) -- (collectPreNodeA\i.east) node {};
      
      \path [draw, ->] (collectPreNodeA\i.east) -- ([yshift=-14 + \i*5.0pt]collectResults.east);
      
      \path [draw, ->] ([yshift=-12 + \i*4.0pt] collectResults.west) to [out=140,in=0] ([yshift=-12 + \i*4.0pt] postprocess0.east);
      \path [draw, ->] ([yshift=-12 + \i*4.0pt] collectResults.west) to [out=180,in=0] ([yshift=-12 + \i*4.0pt] postprocess1.east);
    }

    \path [draw, ->, line width=1.00mm] (postprocess0.west) -- node [above] (preProcessText) {$\ell_0(\{S_i\}_{i=1}^n)$} (updatedState0.east);    
    \path [draw, ->, line width=1.00mm] (postprocess1.west) -- node [above] (preProcessText) {$\ell_1(\{T_i\}_{i=1}^m)$} (updatedState1.east);   
    
    
    \path (collectPreNodeA1.east)+(-3.0, -0.3) node (preProcessText) {$f(S'_{i}) \; | \; g(T'_{i})$};
    
    \path (sampleGen0.east)+(1.0, 0.6) node (preProcessText) {$\{S_i\}_{i=1}^n$};
    \path (sampleGen1.east)+(1.0, 0.6) node (preProcessText) {$\{T_i\}_{i=1}^m$};
    
    \path (preprocess0.east)+(1.0, 0.6) node (preProcessText) {$\{S'_i\}_{i=1}^n$};
    \path (distributeSamples.west)+(-0.4, -1.0) node (preProcessText) {$\{T'_i\}_{i=1}^m$};
    
    \path (collectResults.west)+(0.1, +1.3) node (preProcessText) {$\{f(S'_i)\}_{i=1}^n$};
    \path (postprocess1.east)+(1.0, -0.6) node (preProcessText) {$\{g(T'_i)\}_{i=1}^m$};
 
    \path [draw, line width=0.55mm] ([yshift=20] engine.west |- problem0.north) -- ([yshift=20] conduit.east |- problem0.north) node (frameworkLineNode) {};
    \path ([yshift=5, xshift=-40] problem0.north |- frameworkLineNode) node[align=center,black,anchor=south] (frameworkTitle) {\textbf{Korali Framework}};
    
    \path [draw, line width=0.55mm] ([yshift=20] worker1.west |- problem0.north) --  ([yshift=20] model1.east |- problem0.north) node (scompLineNode) {};
    \path ([yshift=10, xshift=5] rank_1_0.north |- scompLineNode) node (supercomputerTitle) {\textbf{Supercomputing Cluster}};
    
\end{tikzpicture}
\end{adjustbox}
\end{center}
\vspace{-10pt}
\caption{Dataflow of the framework's sampling engine running two experiments, $e_0$ and $e_1$, concurrently. Samples from both experiments are passed through the distribution conduit, which assign them to any available \textit{idle} worker. Here, workers represent teams of \textit{m} MPI ranks, which collaborate to compute an assigned sample. Workers evaluate the model corresponding to the sample's experiment (\textit{f}, for $e_0$, or; \textit{g}, for $e_1$). Upon receiving results, the distribution conduit assigns them to the corresponding experiment.}
    \label{fig:engine_parallel}
\end{figure*}

\subsection{Distributed Sampling Engine}

The sampling engine supports the parallel evaluation of multiple computational models. To enable parallel sampling, the user runs multiple instances of the Korali application, typically via the \textit{mpiexec} command\footnote{For systems that do not support MPI, the \texttt{Concurrent} execution mode can be used to run with multiple concurrent processes using a fork/join strategy instead.}. The engine determines the number $k$ of processes that have been instantiated and assigns roles to them. The first process assumes the role of the \textit{engine}, managing sample production and evaluation, as shown in \cref{fig:engine_simple}. The rest of $k-1$ processes assume the role of \textit{workers}, whose task is to wait for incoming samples, run the computational model $f$, and to return their results.

The distribution of samples and collection of results is managed by a \textit{distribution conduit} module between the experiment and the computational model.  This conduit keeps track of the state of each worker (i.e., \texttt{idle}, \texttt{working}, \texttt{pending}), and assigns incoming samples workers that are in the \texttt{idle} state. As soon as a sample is received, a worker transitions to \texttt{busy} state, during which it executes the model $f$. When the worker finishes executing $f$, it returns the results and sets its state to \texttt{pending}, indicating that the engine can collect the result. The worker state transitions back to \texttt{idle} only after the conduit has retrieved the result. The engine employs an opportunistic strategy for work distribution in which it maintains a common queue of pending samples from which all workers are served. The conduit distributes samples on a one-by-one basis, in real-time, as soon as a worker becomes \texttt{idle}. 

The engine also supports the evaluation of parallel (MPI-based) models. For this case, the execution conduit creates a set of \textit{worker teams}, each assigned a subset of MPI ranks. All ranks from the same team work together in the evaluation of incoming samples. Users define the number of MPI ranks per team ($m$) through the \texttt{"Ranks Per Worker Team"} configuration parameter. For a run with $N$ MPI ranks (as specified in the MPI launch command), the conduit assigns one rank to the sampling engine, and creates $k$ worker teams, each with $\lfloor (N-1)/m \rfloor$ ranks. Every worker team owns their private MPI communicator, which allows message passing between the ranks contained therein. Any MPI-based computational model passed to Korali should use this team-specific communicator for message exchanges. To identify the ranks in a given team, the conduit module appends an \texttt{MPI Com\-mu\-ni\-ca\-tor} field to the sample indicating which group corresponds to the receiving worker. With this value, the model can determine the $m$ number of ranks in the team and the rank identifiers therein. The model can then operate like a regular MPI application and produce a result collaboratively.   

A novelty in the sampling engine is the ability to execute multiple independent experiments concurrently. The goal is to maximize the pool of pending samples at any moment during execution, maximizing worker utilization in the presence of load imbalance (see \cref{subsection:case2}). \Cref{fig:engine_parallel} shows the engine's dataflow when executing two experiments simultaneously ($e_0$, and $e_1$) on a supercomputer cluster. The engine switches its execution context between both experiments, continuously polling whether either is ready to advance to the next generation or return partial results for storage. During execution, each experiment produces and evaluates its own set of samples ($S'$ for $e_0$ and $T'$ for  $e_1$). 

The distribution conduit manages each of the experiment's samples independently, distributing them among the common set of workers. Depending on which experiment has issued the sample, the conduit indicates to the worker which computational model to run. In this case, $f$, if the sample belongs to $e_0$, or; $g$, if the sample belongs to $e_1$. The results are asynchronously returned to the collection module, which distributes them back to the corresponding experiment. The engine evaluates each experiment's termination criteria after the given experiment reaches the end of its current generation.  Experiments advance independently from each other, storing their results in separate folders, and the engine returns when all of them complete.

\subsection{Modularity and Fault-Tolerance} \label{ssec:modules}

\begin{figure}[htb]
\begin{center}
\begin{adjustbox}{trim={1.95cm 0 0 0},clip,width=\columnwidth}
\pgfdeclarelayer{background}
\pgfdeclarelayer{foreground}
\pgfsetlayers{background,main,foreground}

\tikzstyle{subModule}=[draw, fill=blue!20, text width=5em, text centered, minimum height=2.5em]
\tikzstyle{module} = [subModule, text width=6em, fill=black!10, minimum height=4em, rounded corners]
\def\blockdist{2.3}
\def\edgedist{2.5}

\makeatletter
\pgfdeclareshape{document}{
\inheritsavedanchors[from=rectangle] 
\inheritanchorborder[from=rectangle]
\inheritanchor[from=rectangle]{center}
\inheritanchor[from=rectangle]{north}
\inheritanchor[from=rectangle]{south}
\inheritanchor[from=rectangle]{west}
\inheritanchor[from=rectangle]{east}
\backgroundpath{
\southwest \pgf@xa=\pgf@x \pgf@ya=\pgf@y
\northeast \pgf@xb=\pgf@x \pgf@yb=\pgf@y
\pgf@xc=\pgf@xb \advance\pgf@xc by-10pt 
\pgf@yc=\pgf@yb \advance\pgf@yc by-10pt
\pgfpathmoveto{\pgfpoint{\pgf@xa}{\pgf@ya}}
\pgfpathlineto{\pgfpoint{\pgf@xa}{\pgf@yb}}
\pgfpathlineto{\pgfpoint{\pgf@xc}{\pgf@yb}}
\pgfpathlineto{\pgfpoint{\pgf@xb}{\pgf@yc}}
\pgfpathlineto{\pgfpoint{\pgf@xb}{\pgf@ya}}
\pgfpathclose
\pgfpathmoveto{\pgfpoint{\pgf@xc}{\pgf@yb}}
\pgfpathlineto{\pgfpoint{\pgf@xc}{\pgf@yc}}
\pgfpathlineto{\pgfpoint{\pgf@xb}{\pgf@yc}}
\pgfpathlineto{\pgfpoint{\pgf@xc}{\pgf@yc}}
}
}
\makeatother

\tikzstyle{doc}=[%
draw,
thick,
align=center,
color=black,
shape=document,
minimum width=7mm,
minimum height=13.2mm,
shape=document,
inner sep=2ex,
]

\begin{tikzpicture}

    \node[doc] (baseCpp) {Module \\ base .cpp };
    \path (baseCpp.south)+(0,-1.0) node (baseConfig) [doc] {Module \\ .config file};
    \path (baseConfig.south)+(0,-1.0) node (baseHpp) [doc] {Module \\ base .hpp};
    
    \path (baseCpp.north)+(1.3,1) node (a0) {};
    \path (baseHpp.south)+(4.4,-1) node (b0) {};
    \path[fill=yellow!20,rounded corners, draw=black!50, dashed] (a0) rectangle (b0) node {};
    
    \path (a0)+(-0.2,0.2) node[anchor=west] (preprocessorTitle) {\small{\textbf{Source Pre-processor}}};
    
    \path (baseHpp.east)+(0.7,0) node[anchor=west] (declarations) [module] {\footnotesize{Create Attribute Declarations}};
    
    \path (baseCpp.east)+(0.7,0) node[anchor=west] (serialization) [module] {\footnotesize{Create Serialization Routines}};
    
    \path [draw, ->, line width=0.4mm] (baseCpp.east) -- (serialization.west);
    \path [draw, ->, line width=0.4mm] (baseHpp.east) -- (declarations.west);
    
    \path [draw, ->, dashed, line width=0.4mm] (baseConfig.east) to [out=0, in=270] node (curveArrow1) {} (serialization.south);
    \path [draw, ->, dashed, line width=0.4mm] (baseConfig.east) to [out=0, in=90] node (curveArrow2) {} (declarations.north);
      
    \path (serialization.east)+(0.7,0) node[anchor=west] (finalCpp) [doc] {Module \\ final .cpp};
    \path (declarations.east)+(0.7,0) node[anchor=west] (finalHpp) [doc] {Module \\ final .hpp};
    
    \path [draw, ->, line width=0.4mm] (serialization.east) -- (finalCpp.west);
    \path [draw, ->, line width=0.4mm] (declarations.east) -- (finalHpp.west);
    
    \path (finalCpp.north)+(1.5,1) node (a1) {};
    \path (finalHpp.south)+(1.9,-1) node (b1) {};
    \path[rounded corners, draw=black!70] (a1) rectangle (b1);
    
    \path (b1 |- baseConfig)+(-0.23,0) node[anchor=center, rotate=90] (compilerTitle) {\small{\textbf{C++ Compiler}}};
    
    \path [draw, ->, line width=0.4mm] (finalCpp) -- (a1 |- finalCpp);
    \path [draw, ->, line width=0.4mm] (finalHpp) -- (a1 |- finalHpp);
    
    \path (b1 |- baseConfig)+(1.5,0) node (finalObject) [doc] {Module.o};
    \path [draw, ->, line width=0.4mm] (b1 |- baseConfig) -- (finalObject);
    
\end{tikzpicture}
\end{adjustbox}
\end{center}
\caption{Korali's source pre-processor takes the module's \texttt{.cpp} and \texttt{.hpp} files as inputs and appends to them automatically-generated class attributes and serialization routines, based on its \texttt{.config} file. The output source is compiled into an object file and linked to the framework.}
    \label{fig:serialization}
\end{figure}
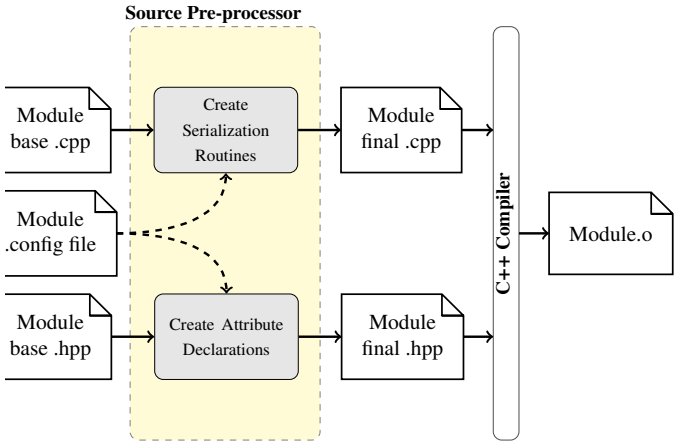

The framework can be extended by adding new problem types and solver modules. To integrate a new module, developers create a new folder in the source code, containing three files: the module's base C++ class declaration header (\texttt{.hpp}) file, its method definitions in the source (\texttt{.cpp}) file, and a configuration (\texttt{.config}) file. Although the module class may contain any arbitrary number of method definitions, it must satisfy a common interface of abstract virtual methods. These methods are necessary to run a generation-based paradigm and depend on whether the new module is of \textit{problem} or \textit{solver} type.

The purpose of the configuration file is to automatize the generation of serialization routines. The engine calls these routines at the end of every generation to save the internal state of a module into a file. The state file serves as a checkpoint from which the execution can be resumed in case of failure or to split large jobs into shorter stints. The engine also stores the internal state of all random number generators to guarantee that the same intermediate and final results are obtained in case of resuming from a file. This approach guarantees reproducible results in long-running experiments that require more than one job to complete.

The engine's source pre-processor creates the serialization routines automatically before compilation, based on the fields declared in the configuration file, as shown in \cref{fig:serialization}. The pre-processor enforces that no class attributes are declared in the header (\texttt{.hpp}) file. Instead, attributes should be inserted as entries in the configuration file specifying, for each one, a field name, a C++ datatype, a description, and a default value. In this way, the framework ensures that the entire internal state of the module is accounted for during serialization. The pre-processor adds these configuration fields as class members to the module's base \texttt{.hpp} header file automatically, so that they can be referenced from the \texttt{.cpp} file. The declaration and definition of the serialization methods are automatically generated and added to the module's source.  A secondary product of code pre-processing is the automatic production of documentation pages for the module's entry in the user manual \cite{kor:manual}.

\section{Experimental Evaluation}
\label{sec:experiments}

We tested Korali on three research studies. First, an optimization study of the shape of fluid transporting pipes. This study shows the use of Korali on a large-scale, high-dimensional optimization job. Second, a hierarchical Bayesian analysis of the dissipation parameter of human \RBC membranes. This study exemplifies the efficiency gains due to scheduling simultaneous experiments in the presence of load imbalance. Lastly, a Bayesian study on the parameters of a coarse-grained water model. This study demonstrates the fault-tolerance mechanisms and reproducibility mechanisms in Korali.

As computational platform, we used both XC40 and XC50 partitions of \pizd \cite{PizDaint2019}, a Cray supercomputer located at the \textit{\CSCS}. The XC40 partition comprises 1'813 compute nodes, each equipped with two Intel Xeon E5-2695-v4 18-core processors running at 2.10GHz and 128 GB RAM. The XC50 partition comprises 5'704 compute nodes, each equipped with a single Intel Xeon E5-2690-v3 12-core processor, running at 2.60GHz and 64GB RAM, and  an NVIDIA ``Tesla'' P100 \GPU with 16GB of device memory. In \ref{appendix:setup}, we provide all resources required to reproduce the results presented in this paper.

\subsection{Study 1: Fluid Transporting Pipes}\label{subsection:case1}

\begin{figure}[htbp]
  \centering
  \includegraphics[scale=0.19]{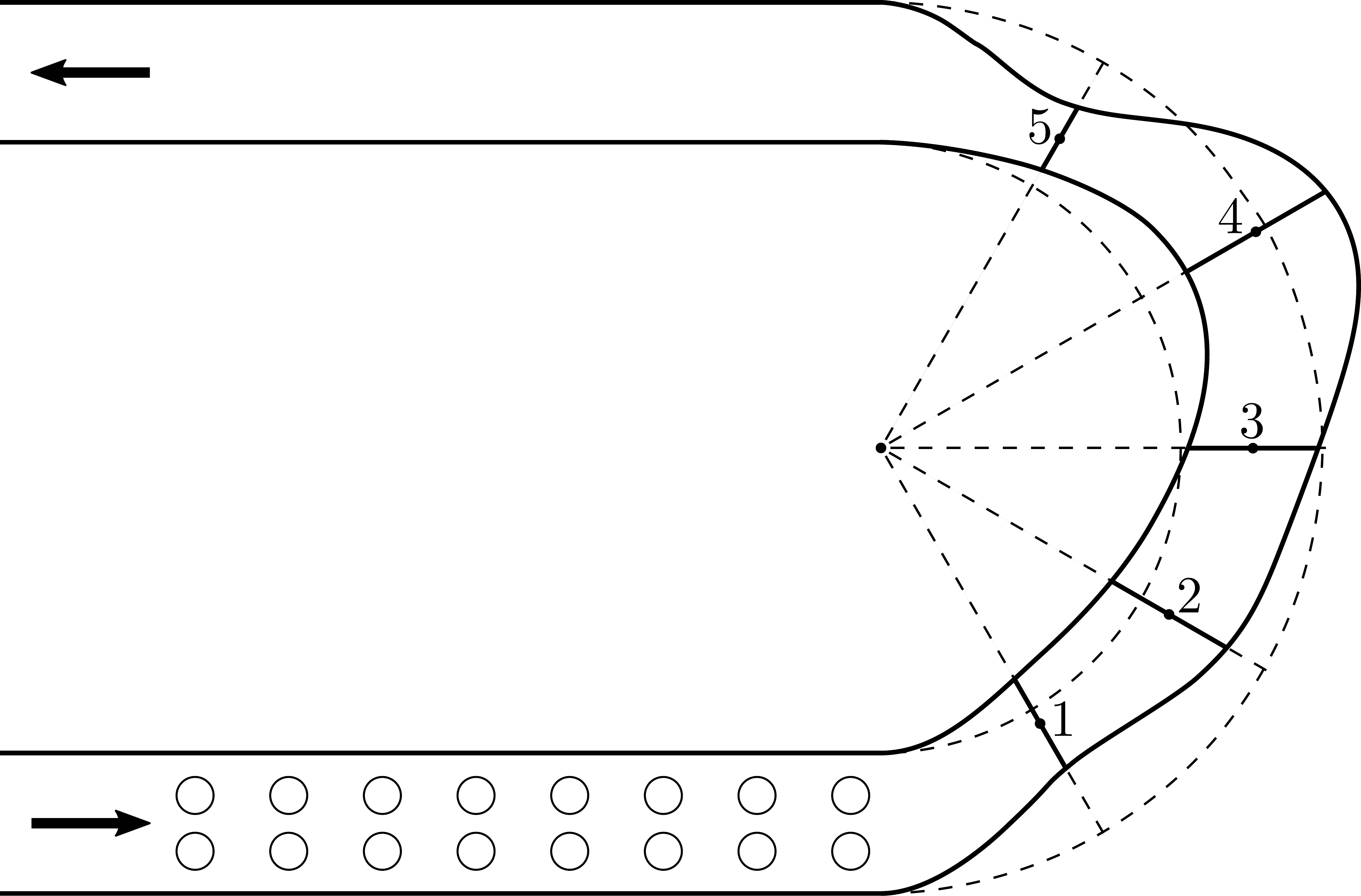}
  \caption{Pipe shape parametrized by width and radial offset
  along five equiangular directions.
  Offsets are relative to the baseline shape (dashed lines).
  Circles show the initial positions of bubbles
  and the arrows indicate the flow direction.}
 \label{fig:pipe_sketch}
\end{figure}

\begin{figure*}[htbp]
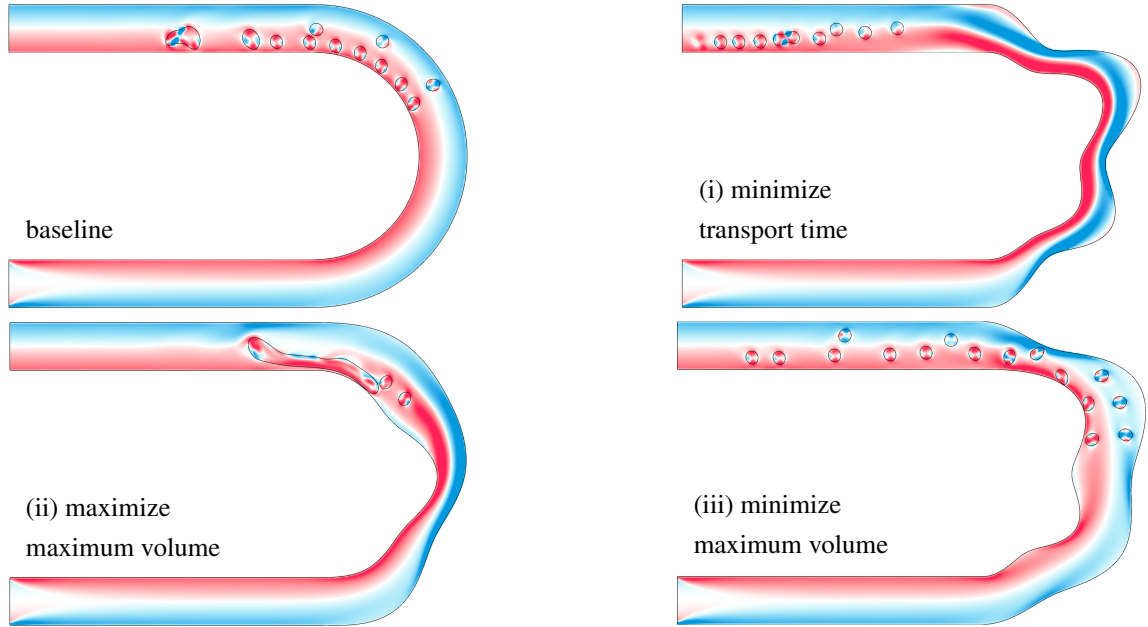

  \newcommand{\p}[3]{%
    \begin{minipage}{\columnwidth}%
    \begin{center}
       \begin{overpic}[unit=1mm,height=4.20cm]{figures/pipe/#1}
         \put(4.5,25){#2}
         \put(4.5,16.5){#3}
       \end{overpic}%
      \end{center}
    \end{minipage}%
  }

  \p{baseline}{}{baseline}%
  \p{MaxMeanVel_gen141_sample000001}{(i) minimize}{transport time}
  \p{MaxNumCoal_gen150_sample000000}{(ii) maximize}{maximum volume}%
  \p{MinNumCoal_gen151_sample000000}{(iii) minimize}{maximum volume}
  \let\p\undefined

  \caption{
    Visualization of the baseline pipe shape and the results for the three optimization cases. The vorticity field is shown in blue, for clockwise, and; red, for counterclockwise. The snapshots are taken at time~$tV/w=14$.
  }
 \label{fig:pipe_opt}
\end{figure*}

Pipe networks are commonly used to convey flowing substances at various scales ranging from microfluidics to industrial facilities. The flow pattern in a pipe is determined by its shape and the flow conditions. Here we apply Korali to optimize the shape of a two-dimensional pipe which transports liquid with bubbles. The pipe consists of two straight segments connected with a U-turn, as illustrated in~\cref{fig:pipe_sketch}, where circulating bubbles can coalesce into larger bubbles. We consider three cases of optimization with different objectives: (i) minimize the transport time of bubbles, \textit{i.e,} the time for all bubbles to exit the pipe; (ii) maximize the maximum bubble volume at the time when it exits the pipe, and; (iii) minimize the maximum bubble volume and therefore achieve a state without coalescence.


The shape of the pipe is defined by 10 parameters that specify the width and radial offset of the U-turn along five directions used as knots of two cubic splines. The flow parameters are the Reynolds number~$\mathrm{Re}=w V\rho_l/\mu_l=100$ and the capillary number~$\mathrm{Ca}=\mu_l V/\sigma=0.1$ defined from the liquid density~$\rho_l$, liquid viscosity~$\mu_l$, surface tension~$\sigma$, pipe width~$w$, and mean inlet velocity~$V$. The gas density and viscosity are set to $0.1\rho_l$ and $0.1\mu_l$.

We find the optimal shape of the pipe for the three cases using \CMAES as optimization algorithm, with a population size of 32 samples per generation. As computational model, we used a finite-volume solver (\aphros \cite{aphros,karnakov2020aphros}) with embedded boundaries \cite{colella2006} to treat complex geometries and a particle method \cite{karnakov2020hybrid} for surface tension forces. Each instance of \aphros ran on 4 nodes (72 cores) of the XC40 partition, a total of 128 nodes (2'304 cores) per case. Each case ran for an average of 15 hours and consumed approximately 2k node hours.

Results of the optimization are shown in~\cref{fig:pipe_opt}. Case (i) results in a shape where the U-turn contracts and then expands to redirect the bubbles towards the centerline, where velocity is maximized. The transport time of the optimized shape has decreased by a factor of~2.1 compared to the baseline.  Case (ii) forms a cavity at the end of the U-turn where the flow stagnates and all bubbles except one coalesce into one elongated bubble. Finally, case (iii) results in a wide shape where bubbles circulate in parallel lanes, achieving a state without coalescence.

\subsection{Study 2: Red Blood Cell Membrane Viscosity}\label{subsection:case2}

\begin{figure}[htbp]
\centering

 \includegraphics[trim=3 608 1300 3,clip,width=0.87\columnwidth]{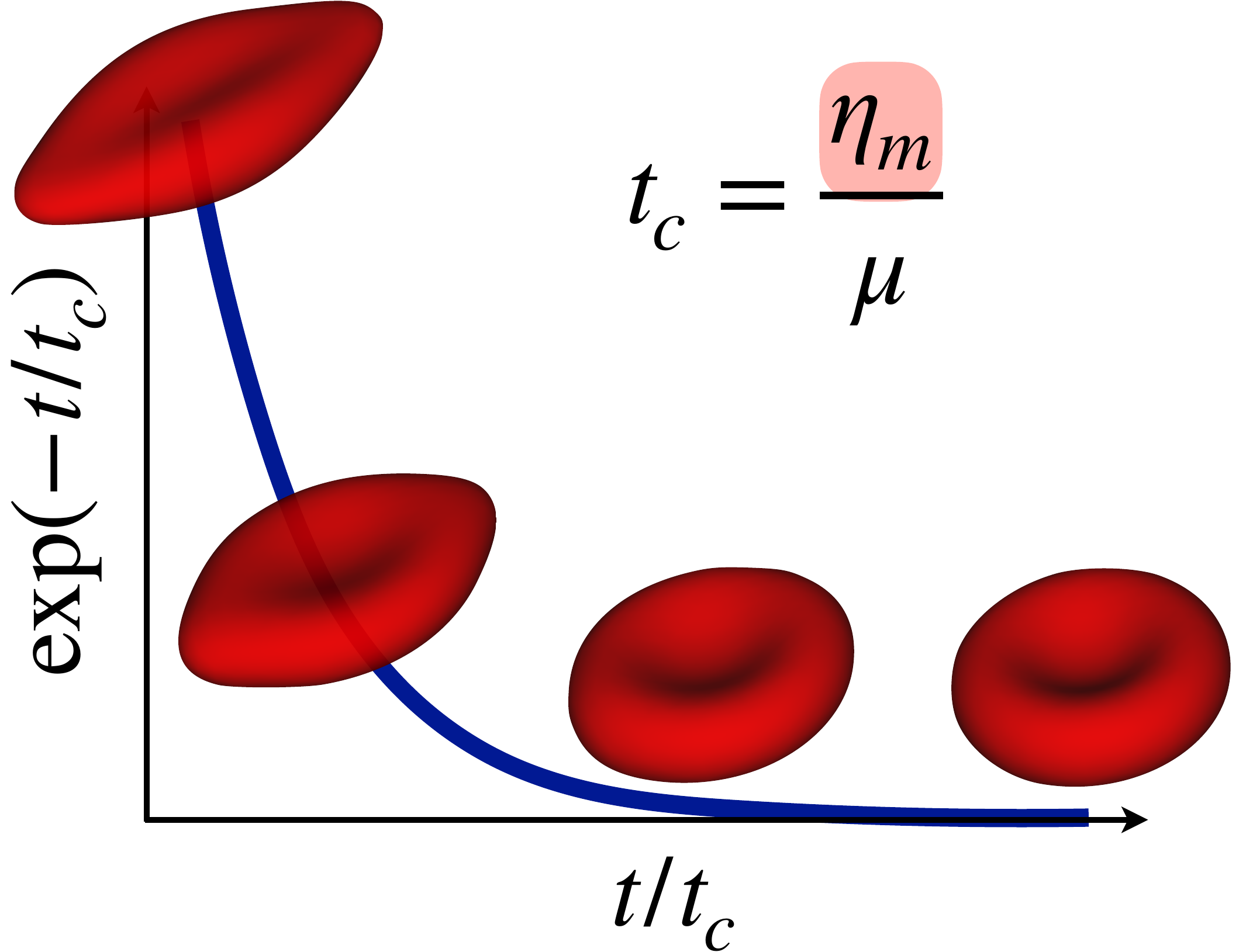}
 \caption{
 Relaxation of a \RBC from an initially elongated state to the biconcave resting shape.
 The time evolution of the length and width of the \RBC are non-dimensionalized following \cite{Hochmuth1979}, with the dimensionless \RBC size following an exponential decay.
 Time is non-dimensionalized by the characteristic relaxation time, $t_c=\eta_m/\mu$, with $\mu$ the elastic shear modulus and $\eta_m$ the \RBC (2D) membrane viscosity. The latter can be inferred given a set of experimental measurements \cite{Waelchli2020}.
 }
 \label{fig:rbc_setup_relax}

\end{figure}

\begin{figure*}[htbp]
 \includegraphics[width=\textwidth, trim={0 0.7cm 0 0},clip]{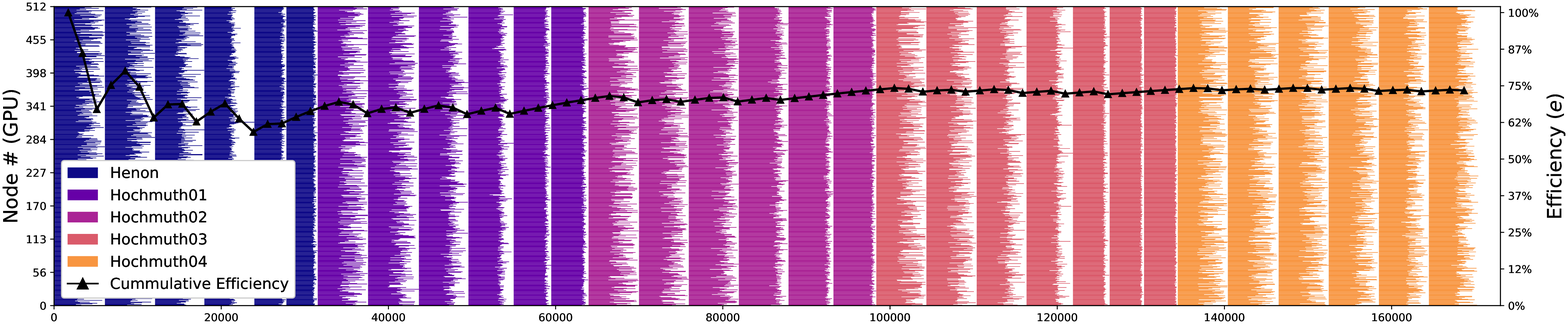}
 \includegraphics[width=\textwidth]{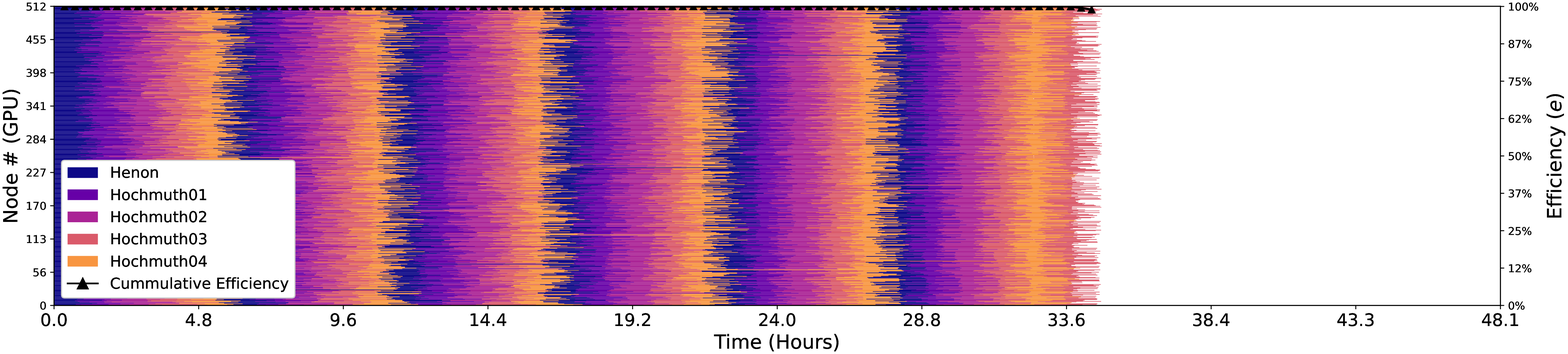}
     \caption{Core usage timelines of Korali during sampling of the experimental datasets, starting from Henon (darkest shade), Hochmuth01, Hochmuth02, Hochmuth03, and ending on the right with Hochmuth04 (lightest shade). The figure shows two timelines: on top, with sequentially scheduled \BASIS experiments, and; on bottom: multiple experiments scheduled simultaneously. The horizontal axis represents the elapsed time (minutes) from the start of the experiment. On the vertical axis, each line represents a different node. Solid lines represent the execution of the model. Blank spaces represent times where a node is idle. The black line indicates the cumulative sampling efficiency ($e$) across time. A higher efficiency reflects a better node usage.}
    \label{fig:timelines_hierarchical}
\end{figure*}

\begin{table}[htbp]
\centering
\footnotesize
\begin{tabular}{lc|c|c|c|cc}
\cline{3-4}
& \multicolumn{1}{l}{}  & \multicolumn{2}{|c|}{\textbf{Node Hours}} & \multicolumn{1}{c}{\textbf{}}  \\ \hline
\multicolumn{1}{|c|}{\textbf{Scheduling}}             & \textbf{Time}    & \textbf{Total} & \textbf{Idle} & \multicolumn{1}{c|}{\textbf{$e$}} & \multicolumn{1}{c|}{\textbf{Energy}} \\ \hline
\multicolumn{1}{|l|}{\textbf{Single}}  & 47.32 $\si{\hour}$               & 24227                & 6604            & \multicolumn{1}{c|}{\textbf{72.7\%}}     & \multicolumn{1}{c|}{10.45 $\si{\giga\joule}$}                           \\ \hline
\multicolumn{1}{|l|}{\textbf{Multiple}}      & 34.78 $\si{\hour}$                & 17809                 & 186             & \multicolumn{1}{c|}{\textbf{98.9\%}}     & \multicolumn{1}{c|}{7.80 $\si{\giga\joule}$}                            \\ \hline
\end{tabular}
\caption{Performance comparison between the two scheduling strategies employed in the \RBC stretching experiment: Single, with individually scheduled experiments, and; Multiple, with all experiments scheduled simultaneously. Here, $e$ represents sampling efficiency. The energy usage measurements were obtained from \pizd's job scheduler.}
\label{tab:results}
\end{table}

\Acp{RBC} are highly deformable objects that incur complex dynamical transitions in response to external disturbances. These transitions lay the foundation for understanding the rheology of blood in flows through capillaries, vascular networks, and medical devices. There is still significant uncertainty in the choice of the mechanical law to describe the \RBC properties, as well as in the parameter values of each model \cite{siguenza2017should}.

In this study, we infer the membrane viscosity which controls the relaxation time of an \RBC membrane. Here, the \RBC membrane is modeled as a collection of particles placed on the nodes of a triangular network \cite{Fedosov2010f}. We used data from five experimental observations (four from \texttt{Hochmuth} \cite{Hochmuth1979} and one from \texttt{Henon} \cite{Henon1999}), on the relaxation of a stretched \RBC to its equilibrium shape in order to infer the posterior distribution of the membrane viscosity ($\eta_{m}$, see \cref{fig:rbc_setup_relax}), and its associated uncertainty ($\sigma$). Due to the presence of heterogeneous data, we employed a hierarchical Bayesian model. The sampling of the posterior distribution is approximated by a two stage importance sampling algorithm. A detailed description of the statistical model, the experimental data, and the results of the hierarchical model can be found in a previous work \cite{Waelchli2020}.

Here, we analyze the performance of the framework during the first stage, where the parameters were sampled individually, conditioned on each experimental data set. For sampling, we employed \BASIS, a reduced bias variant of the \TMCMC algorithm for the sampling of the posterior distribution. \BASIS is a sampling algorithm tailored for Bayesian uncertainty quantification and targeted to parallel architectures \cite{Wu2016}. We configured \BASIS to run a population size of 512 samples per generation. For the computational model, we used \mir \cite{alexeev2019mirheo}, a high-performance GPU-based library for microfluidic simulations. 

We ran the five experiments on 512 nodes of the XC40 partition using 512 MPI ranks, each running an instance of \mir per node. In a previous work \cite{Waelchli2020}, we had found that the \RBC membrane relaxation model shows a high variance in running times ($40\sim100$min per sample). This variance caused a workload imbalance among the workers, with detrimental impact on the performance of the \BASIS algorithm. The effect can be appreciated when running each of the five \BASIS sampling experiments individually, as shown in \cref{fig:timelines_hierarchical} (top). The five experiments took 48.1 hours to complete on 512 nodes, requiring a total of 24.6k node hours. This approach yielded sampling efficiency\footnote{Calculated as the ratio between busy and idle time.} ($e$) of $72.7\%$. It follows that nodes remained idle 27.3\% of their running time. \Cref{tab:results} (row: \textit{Single}) shows that only this resulted in a loss of 6.6k (idle) node hours. In total, the energy usage, as reported from \pizd's job scheduler, was of 10.45 \si{\giga\joule}.

To alleviate the effect of load imbalance, we configured Korali to schedule all five experiments simultaneously. The timeline in \Cref{fig:timelines_hierarchical} (bottom) shows that nodes remained busy during most of the run with the multi-experiment variant. The results, summarized in \cref{tab:results}, indicate that this approach yields a superior efficiency (98.9\%) compared to the former approach, wasting much fewer node hours (186), as well as requiring less energy (7.80 \si{\giga\joule}). Furthermore, it also reduced the run-to-completion time from 47.32 to 34.78 hours. 

\subsection{Study 3: Coarse-Grained Water Model}\label{subsection:case3}

\begin{figure}[htbp]
\centering
 \includegraphics[width=\columnwidth]{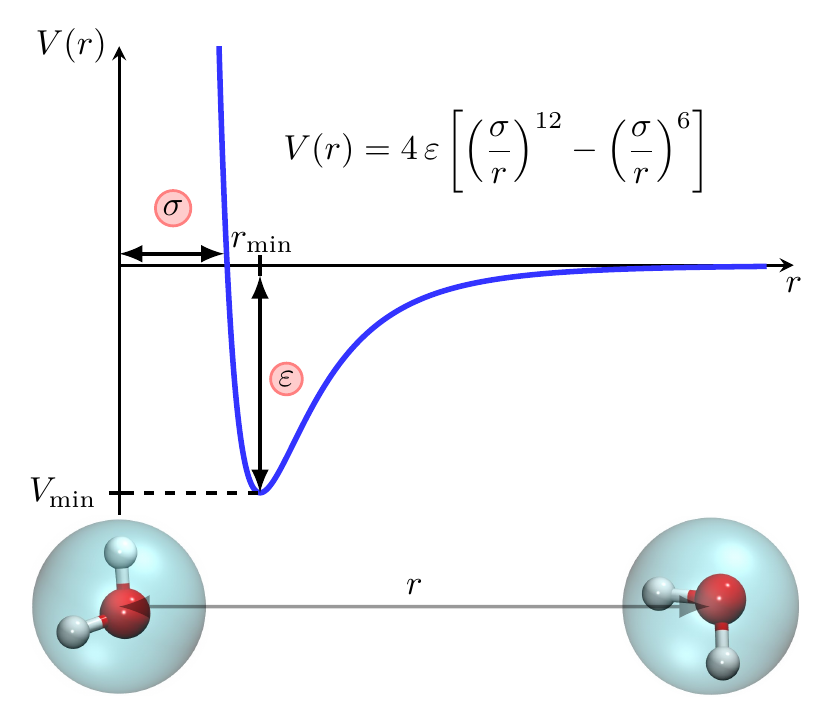}
 \caption{The Lennard-Jones potential is a function of the distance ($r$) between two water molecules, represented as solid spheres. It is repulsive for distances less than $r_\textrm{min}$ and attractive for larger distances.
 The potential is parameterized by $\varepsilon$ that controls the depth of the well ($V_\textrm{min}$), and by $\sigma$ that controls the change from repulsion to attraction point.}
    \label{fig:lennardJonesPotential}
\end{figure}

In this study, we apply Bayesian inference on the assumed Lennard-Jones potential between particles with two parameters ($\varepsilon$ and $\sigma$) for a coarse-grained water model where a single water molecule is represented by a solid sphere (see \cref{fig:lennardJonesPotential}). We reproduced the computational setup of a previous work \cite{Zavadlav2019}, where the parameters of the model are calibrated to experimental data (density, dielectric constant, surface tension, isothermal compressibility, and shear viscosity) at fixed temperature and pressure. To find the parameters that maximize the posterior distribution of the parameters, we use \CMAES as optimization algorithm, with a population size of 16 samples per generation. For the computational model, we used \lammps (Large-scale Atomic/Molecular Massively Parallel Simulator) \cite{lammps}, a well-known molecular dynamics simulation library that models atoms or ensembles of particles in solid, liquid or gaseous state. We run the experiment using 16 compute nodes of the XC50 partition and two workers per node. Each worker ran an instance of \lammps using 2 MPI ranks over 12 OpenMP threads. 

\begin{figure}[t]
\centering
 \includegraphics[width=\columnwidth]{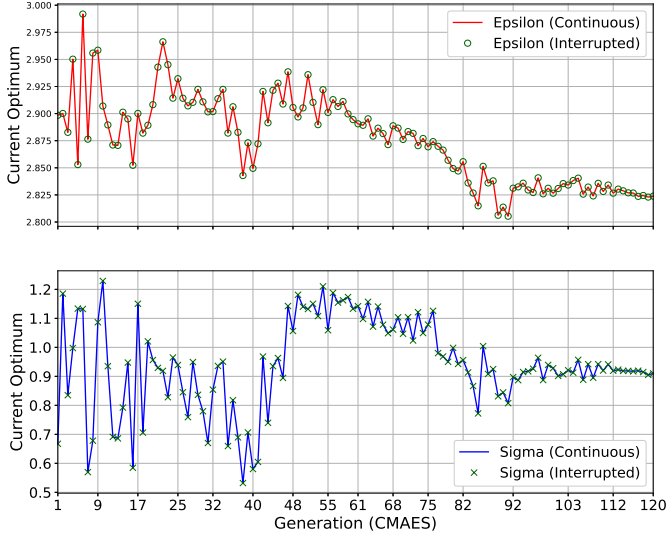}
     \caption{Evolution of the parameter values that maximize the posterior distribution of the Bayesian problem, showing: in solid lines, the run in which  \CMAES was uninterrupted, and; in markers, the run which was interrupted every 15 minutes. Vertical grid lines indicate the generations at which the interrupted run was restarted.}
    \label{fig:comparison_restart}
\end{figure}

Here, we validate the framework's fault-tolerance and reproducibility by running the same study twice. For the first run, we allow Korali to complete without interruptions. For the second run, we allow it to run for only 15 minutes at a time before forcing the job scheduler to terminate it. To reach the final results with the interrupted run, we re-schedule its launcher job after each interruption, reloading the internal state of \CMAES upon restart. Since both runs use the same random seed, we expect to observe the exact same intermediate and final results. \Cref{fig:comparison_restart} shows the comparison of the per-generation evolution of parameter optima between the single-run execution (continuous line), and the interrupted execution (markers) for the two optimization parameters. In the figure, vertical grid lines indicate the generation at which the interruptions occurred for the latter, for a total of 16 restarts. Results show that the optimal parameters and their convergence path was identical for both runs. 

\section{Related Work} \label{section:related}

\begin{figure*}[htbp]
\centering

 \subfloat{{\includegraphics[width=\columnwidth]{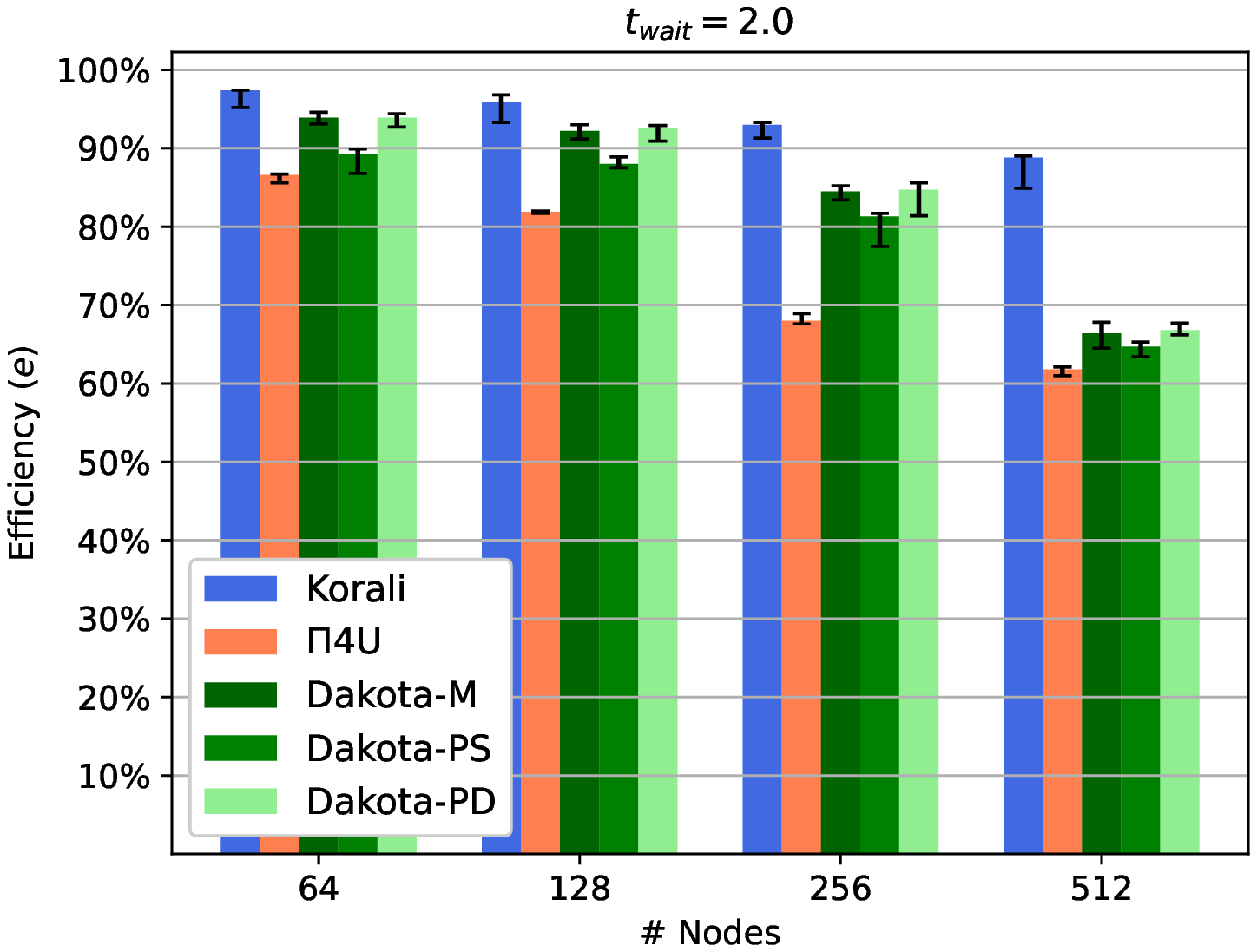} }} \subfloat{{\includegraphics[width=\columnwidth]{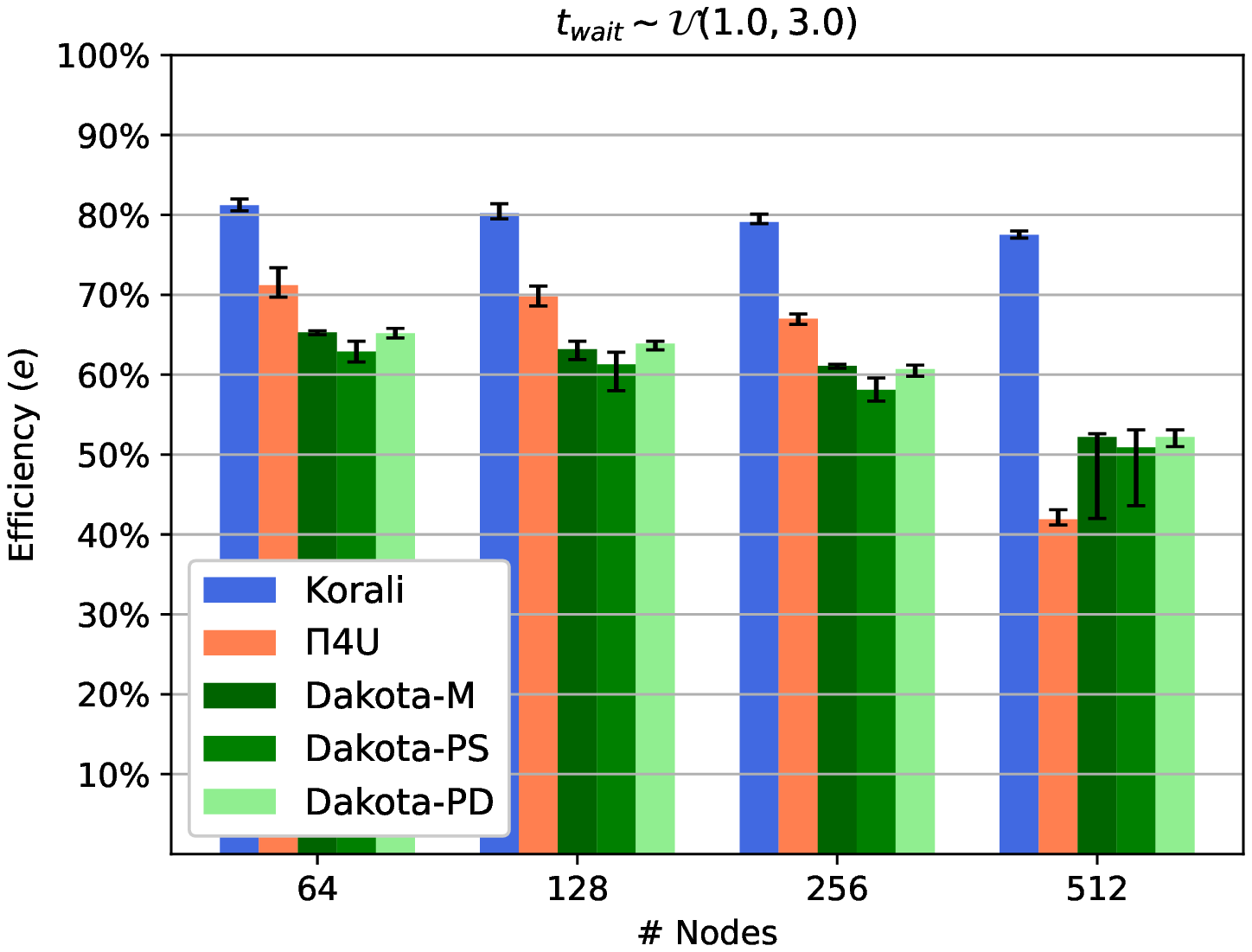} }}

 \caption{Weak scaling studies comparing the sampling efficiency ($e$) of Korali, $\Pi$4U, and Dakota (three variants) for a synthetic benchmark (left) without ($T_{wait} = 2.0$), and (right) with load imbalance ($T_{wait} \sim \mathcal{U}(1.0, 3.0)$) on 64, 128, 256, and 512 nodes. The colored bars highlight the median efficiencies, and the black intervals indicate the maximum and minimum.}
\label{fig:benchs}
\end{figure*}

Many of the problems and solver methods currently implemented in Korali can also be recognized across other statistical \UQ softwares.  These include \textit{ABC-SysBio} \cite{abcsysbio2010}, \textit{APT-MCMC} \cite{aptmcmc, aptmcmc:page}, \textit{BCM} \cite{bcm, bcm:page}, \textit{BioBayes} \cite{biobayes2008}, \textit{Dakota} \cite{dalbey2020dakota}, \textit{EasyVVUQ} \cite{richardson2020easyvvuq}, \textit{$\Pi$4U} \cite{Hadjidoukas2015}, \textit{MUQ} \cite{muq:page}, \textit{PSUADE}  \cite{psuade:page}, \textit{QUESO} \cite{queso2011}, \textit{ScannerBit} \cite{scannerbit} (a \textit{GAMBIT} \cite{gambit, gambit:page}  module), which are standalone applications; \textit{Chaospy} \cite{chaospy:page}, \textit{Uncertainpy} \cite{uncertainpy, uncertainpy:page},
\textit{UQ Toolkit} \cite{ghanem2017}, \textit{UQPy} \cite{uqpy:page}, which are publicly available Python packages; \textit{Stan} \cite{stan:page}, a programming language for statistical inference;  and \textit{UQLab} \cite{uqlab:page}, a MATLAB framework. 

Here, we analyze in further detail the $\Pi$4U and Dakota frameworks as they offer support for distributed sampling and thus relate closer to Korali's goal. The {$\Pi$4U} \cite{Hadjidoukas2015} framework is one of the first efforts in providing support of distributed UQ experiments and, to the best of our knowledge, the only to have reported detailed performance metrics at scale. $\Pi$4U employs the TORC tasking library \cite{Hadjidoukas2012} to distribute the execution of samples to workers.  \textit{Dakota} \cite{dalbey2020dakota} is a well-established C++-based framework for uncertainty quantification, that interfaces with simulation software through a multi-level MPI-based parallelism interface. Dakota is used in a wide-range of applications for the US Department of Energy \cite{adams2018dakota}. 

To compare the efficiency between frameworks, we created a synthetic benchmark that runs a single-variable optimization experiment on a computational model that passively waits for a given number of seconds ($T_{wait}$) before returning a random result. By employing a passive wait, we can fix the running time of each sample, ruling out time variances common to compute-intensive computational models. To drive sampling, we employ evolutionary optimization algorithms ({CMAES} \cite{Hansen2016}, for Korali and $\Pi$4U, and; {COLIN-EA} \cite{Hart2003}, for Dakota) configured to run 5 generations, $4N$ samples per generation\footnote{For fairness, we verified that the generation pre- and post-processing times for all algorithms are negligible.}, where $N$ is the number of workers, and one node per worker. This configuration allows to conduct weak scaling studies, evaluating the impact of node scaling on efficiency, while keeping the workload per worker constant. We configured average sample times to add up to 40 seconds in total per experiment, which represents their ideal running time. We measure efficiency ($e$) for each framework by dividing the ideal time by its actual running time.

We test the following 5 variants: Korali, $\Pi$4U, Dakota-M (master/worker scheduler), Dakota-PS (static scheduler), and Dakota-PD (dynamic scheduler). We use the synthetic benchmark to run two weak scaling studies, with and without load imbalance. To account for the effect of stochastic waiting times, we ran 10 repetitions of each experiment. In \ref{appendix:setup}, we provide the experimental setup and data necessary to replicate the results.

The first study represents a scenario where there is no load imbalance ($T_{wait} = 2.0$ seconds, for all samples). Here, we measure the inherent efficiency of the frameworks in distributing samples to workers without the detrimental effect of imbalance. The gap between the attained and an ideal efficiency therefore illustrates the time spent on communication, I/O operations, and scheduling overhead only.  \cref{fig:benchs} (left) shows the results of weak scaling by running the 5 variants from 64 to 512 nodes of the Piz Daint supercomputer. All variants provide high efficiencies ($86\% < e < 97\%$) at smaller scales (64 and 128 nodes), with Korali as the most efficient by a small margin. At the largest scale (512 nodes), the  differences are evident, since both $\Pi$4U ($e = 62\%$) and Dakota ($e = 67\%$) appear to be especially susceptible to the increasing scheduling costs, compared to Korali ($e = 86\%$). 

The second study simulates experiments with a high load imbalance. Here, the waiting time for each sample is drawn from a random variable $T_{wait} \sim \mathcal{U}(1.0, 3.0)$ seconds. We consider the same ideal case (in average, 40 seconds per experiment) as basis for the calculation of efficiency. \cref{fig:benchs} (right) show the results for this study. We observe that load imbalance plays a detrimental effect on the efficiency of all variants. However, Korali is the least affected of them throughout all scales. We observe the larger difference between the variants when running on 512 nodes and the larger imbalance, where $\Pi$4U and Dakota show a low efficiency ($e = 41\%$ and $e = 52\%$, respectively), while Korali sustains a higher performance ($e = 78\%$).

\section{Conclusions and Future Work}\label{sec:conclusion}

We introduced Korali, a unifying framework for the efficient deployment of Bayesian \UQ and optimization methods on large-scale supercomputing systems. The software enables a scalable, fault-tolerant, and reproducible execution of experiments from a wide range of problem types and solver methods. The experimental results have shown that Korali is capable of attaining high sampling efficiencies on up to 512 nodes of \pizd. Furthermore, we have shown that running multiple experiments simultaneously on a common set of computational resources minimizes load imbalance and achieves almost perfect node usage, even in the presence of high sample runtime variance. We have also shown that the framework can run out-of-the-box libraries, such as \lammps, while providing fault tolerance mechanisms. Finally, we demonstrate that Korali attains higher efficiencies than other prominent distributed sampling frameworks.

We are currently integrating support for distributed \RL methods \cite{nair2015massively, espeholt2018} that target the incremental optimization of a policy for multiple concurrent agents exploring and acting in a virtual environment to collect and maximize rewards. The architecture for these methods closely resembles the workflow provided by Korali, thus making them suitable candidates for integration. We believe that a platform integrating stochastic optimization, \UQ and \RL will be of great value for the broad scientific community. 

\section*{Acknowledgements}
We would like to thank L. Amoudruz, I. Kicic, P. Weber, and  F. Wermelinger  for providing us with their support and invaluable feedback on Korali's design and the writing of this article. We acknowledge support by the European Research Council (ERC Advanced Grant 341117) and computational resources granted by the Swiss National Supercomputing Center (CSCS) under project ID s929.

\bibliographystyle{elsarticle-num}
\bibliography{references}

\begin{thebibliography}{10}
\expandafter\ifx\csname url\endcsname\relax
  \def\url#1{\texttt{#1}}\fi
\expandafter\ifx\csname urlprefix\endcsname\relax\def\urlprefix{URL }\fi
\expandafter\ifx\csname href\endcsname\relax
  \def\href#1#2{#2} \def\path#1{#1}\fi

\bibitem{ashby:2010}
S.~Ashby, P.~Beckman, J.~Chen, P.~Colella, B.~Collins, D.~Crawford,
  J.~Dongarra, D.~Kothe, R.~Lusk, P.~Messina, et~al., {The opportunities and
  challenges of exascale computing}, Summary Report of the Advanced Scientific
  Computing Advisory Committee (ASCAC) Subcommittee (2010) 1--77.

\bibitem{Hansen2003}
N.~Hansen, S.~D. Muller, P.~Koumoutsakos, {Reducing the Time Complexity of the
  Derandomized Evolution Strategy with Covariance Matrix Adaptation (CMA-ES)
  Nikolaus}, Evolutionary Computation 11~(1) (2003) 1--18.

\bibitem{Kern2004}
S.~Kern, S.~D. M{\"u}ller, N.~Hansen, D.~B{\"u}che, J.~Ocenasek,
  P.~Koumoutsakos, Learning probability distributions in continuous
  evolutionary algorithms --a comparative review, Natural Computing 3~(1)
  (2004) 77--112.

\bibitem{Akimoto:2014}
Y.~Akimoto, A.~Auger, N.~Hansen,
  \href{https://hal.inria.fr/hal-00997835}{{Comparison-Based Natural Gradient
  Optimization in High Dimension}}, in: {Genetic and Evolutionary Computation
  Conference GECCO'14}, {ACM}, Vancouver, Canada, 2014.
\newline\urlprefix\url{https://hal.inria.fr/hal-00997835}

\bibitem{uqlab:page}
{UQLab}, \url{https://www.uqlab.com}, (2019-06-13) ({}).

\bibitem{easyvvuq:page}
{EasyVVUQ}, \url{https://easyvvuq.readthedocs.io/en/latest/}, (2019-06-13)
  ({}).

\bibitem{abcsysbio2010}
J.~Liepe, C.~Barnes, E.~Cule, K.~Erguler, P.~Kirk, T.~Toni, M.~P. Stumpf,
  {ABC-SysBio—approximate Bayesian computation in Python with GPU support},
  Bioinformatics 26~(14) (2010) 1797--1799.

\bibitem{aptmcmc}
L.~A. Zhang, A.~Urbano, G.~Clermont, D.~Swigon, I.~Banerjee, R.~S. Parker,
  {APT-MCMC, a C++/Python implementation of Markov Chain Monte Carlo for
  parameter identification}, Computers \& Chemical Engineering 110 (2018).

\bibitem{psuade:page}
{PSUADE},
  \url{https://computation.llnl.gov/projects/psuade-uncertainty-quantification},
  (2019-06-13) ({}).

\bibitem{queso2011}
E.~E. Prudencio, K.~W. Schulz, {The Parallel C++ Statistical Library `QUESO':
  Quantification of Uncertainty for Estimation, Simulation and Optimization},
  in: Euro-Par 2011: Parallel Processing Workshops, Springer Berlin Heidelberg,
  Berlin, Heidelberg, 2012, pp. 398--407.

\bibitem{stan:page}
{Stan}, \url{https://mc-stan.org}, (2019-06-13) ({}).

\bibitem{Hadjidoukas2015}
P.~Hadjidoukas, P.~Angelikopoulos, C.~Papadimitriou, P.~Koumoutsakos, {$\Pi$4U:
  A high performance computing framework for Bayesian uncertainty
  quantification of complex models}, Journal of Computational Physics 284
  (2015) 1--21.

\bibitem{dalbey2020dakota}
K.~Dalbey, M.~S. Eldred, G.~Geraci, J.~D. Jakeman, K.~A. Maupin, J.~A.
  Monschke, D.~T. Seidl, L.~P. Swiler, A.~Tran, F.~Menhorn, et~al., {Dakota A
  Multilevel Parallel Object-Oriented Framework for Design Optimization
  Parameter Estimation Uncertainty Quantification and Sensitivity Analysis:
  Version 6.12 Theory Manual.}, Tech. rep., Sandia National Lab (2020).

\bibitem{dongarra:Exascale}
J.~Dongarra, P.~Beckman, T.~Moore, P.~Aerts, G.~Aloisio, J.-C. Andre, et~al.,
  {The International Exascale Software Project Roadmap}, Int. J. High Perform.
  Comput. Appl. 25~(1) (2011) 3--60.

\bibitem{Adam2015}
D.~P. Kingma, J.~Ba, \href{http://arxiv.org/abs/1412.6980}{Adam: {A} method for
  stochastic optimization}, in: Y.~Bengio, Y.~LeCun (Eds.), 3rd International
  Conference on Learning Representations, {ICLR} 2015, San Diego, CA, USA, May
  7-9, 2015, Conference Track Proceedings, 2015.
\newline\urlprefix\url{http://arxiv.org/abs/1412.6980}

\bibitem{Gelman2014}
M.~D. Homan, A.~Gelman, The no-u-turn sampler: Adaptively setting path lengths
  in hamiltonian monte carlo, J. Mach. Learn. Res. 15~(1) (2014) 1593–1623.

\bibitem{feroz2009}
F.~Feroz, M.~P. Hobson, M.~Bridges,
  \href{https://doi.org/10.1111/j.1365-2966.2009.14548.x}{{MultiNest: an
  efficient and robust Bayesian inference tool for cosmology and particle
  physics}}, Monthly Notices of the Royal Astronomical Society 398~(4) (2009)
  1601--1614.
\newblock \href
  {http://arxiv.org/abs/https://academic.oup.com/mnras/article-pdf/398/4/1601/3039078/mnras0398-1601.pdf}
  {\path{arXiv:https://academic.oup.com/mnras/article-pdf/398/4/1601/3039078/mnras0398-1601.pdf}},
  \href {https://doi.org/10.1111/j.1365-2966.2009.14548.x}
  {\path{doi:10.1111/j.1365-2966.2009.14548.x}}.
\newline\urlprefix\url{https://doi.org/10.1111/j.1365-2966.2009.14548.x}

\bibitem{Wu2018}
S.~Wu, P.~Angelikopoulos, C.~Papadimitriou, P.~Koumoutsakos, {Bayesian Annealed
  Sequential Importance Sampling (BASIS): an unbiased version of Transitional
  Markov Chain Monte Carlo}, ASCE-ASME Journal of Risk and Uncertainty in
  Engineering Systems, Part B: Mechanical Engineering 4~(1) (2018).

\bibitem{kor:manual}
{Korali User Manual}, \url{https://www.cse-lab.ethz.ch/korali/docs/},
  (2021-01-15) ({}).

\bibitem{PizDaint2019}
{CSCS Piz Daint}, \url{https://www.cscs.ch/computers/piz-daint/}, (30-10-2019)
  ({}).

\bibitem{aphros}
\relax{Aphros: Parallel solver for incompressible multiphase flows},
  \url{https://github.com/cselab/aphros} (2020).

\bibitem{karnakov2020aphros}
P.~Karnakov, F.~Wermelinger, S.~Litvinov, P.~Koumoutsakos, Aphros: High
  performance software for multiphase flows with large scale bubble and drop
  clusters, in: Proceedings of the Platform for Advanced Scientific Computing
  Conference, 2020, pp. 1--10.

\bibitem{colella2006}
P.~Colella, D.~T. Graves, B.~J. Keen, D.~Modiano, A cartesian grid embedded
  boundary method for hyperbolic conservation laws, Journal of Computational
  Physics 211~(1) (2006) 347--366.

\bibitem{karnakov2020hybrid}
P.~Karnakov, S.~Litvinov, P.~Koumoutsakos, A hybrid particle volume-of-fluid
  method for curvature estimation in multiphase flows, International Journal of
  Multiphase Flow 125 (2020) 103209.

\bibitem{Hochmuth1979}
R.~M. Hochmuth, P.~Worthy, E.~A. Evans, Red cell extensional recovery and the
  determination of membrane viscosity, Biophysical journal 26~(1) (1979)
  101--114.

\bibitem{Waelchli2020}
D.~W\"{a}lchli, S.~M. Martin, A.~Economides, L.~Amoudruz, G.~Arampatzis,
  X.~Bian, P.~Koumoutsakos, {Load Balancing in Large Scale Bayesian Inference},
  in: Proceedings of the Platform for Advanced Scientific Computing Conference,
  PASC '20, ACM, 2020, pp. 1--12.

\bibitem{siguenza2017should}
J.~Sig{\"u}enza, S.~Mendez, F.~Nicoud, {How should the optical tweezers
  experiment be used to characterize the red blood cell membrane mechanics?},
  Biomechanics and modeling in mechanobiology 16~(5) (2017) 1645--1657.

\bibitem{Fedosov2010f}
D.~A. Fedosov, B.~Caswell, G.~E. Karniadakis, {A Multiscale Red Blood Cell
  Model with Accurate Mechanics, Rheology, and Dynamics}, Biophysical Journal
  98~(10) (2010) 2215--2225.

\bibitem{Henon1999}
S.~H{\'{e}}non, G.~Lenormand, A.~Richert, F.~Gallet, {A new determination of
  the shear modulus of the human erythrocyte membrane using optical tweezers},
  Biophysical Journal 76~(2) (1999) 1145--1151.

\bibitem{Wu2016}
S.~Wu, P.~Angelikopoulos, G.~Tauriello, C.~Papadimitriou, P.~Koumoutsakos,
  {Fusing heterogeneous data for the calibration of molecular dynamics force
  fields using hierarchical Bayesian models}, The Journal of Chemical Physics
  145~(24) (2016).

\bibitem{alexeev2019mirheo}
D.~Alexeev, L.~Amoudruz, S.~Litvinov, P.~Koumoutsakos, Mirheo: High-performance
  mesoscale simulations for microfluidics (2020).

\bibitem{Zavadlav2019}
J.~Zavadlav, G.~Arampatzis, P.~Koumoutsakos, {Bayesian selection for
  coarse-grained models of liquid water}, Scientific Reports 9~(1) (2019)
  1--10.

\bibitem{lammps}
S.~Plimpton, {Fast Parallel Algorithms for Short – Range Molecular Dynamics},
  Journal of Computational Physics 117 (1995) 1--19.

\bibitem{aptmcmc:page}
{APT-MCMC}, \url{https://apt-mcmc.readthedocs.io/en/latest/}, (2020-02-27)
  ({}).

\bibitem{bcm}
B.~Thijssen, T.~M.~H. Dijkstra, T.~Heskes, L.~F.~A. Wessels, Bcm: toolkit for
  bayesian analysis of computational models using samplers, BMC Systems Biology
  10~(1) (2016) 100.

\bibitem{bcm:page}
{BCM}, \url{http://ccb.nki.nl/software/bcm/}, (2019-06-14) ({}).

\bibitem{biobayes2008}
V.~Vyshemirsky, M.~Girolami, {BioBayes: A software package for Bayesian
  inference in systems biology}, Bioinformatics 24~(17) (2008) 1933--1934.

\bibitem{richardson2020easyvvuq}
R.~A. Richardson, D.~W. Wright, W.~Edeling, V.~Jancauskas, J.~Lakhlili, P.~V.
  Coveney, Easyvvuq: A library for verification, validation and uncertainty
  quantification in high performance computing, Journal of Open Research
  Software 8~(1) (2020).

\bibitem{muq:page}
{MUQ - MIT Uncertainty Quantification Library}, \url{http://muq.mit.edu},
  (2020-02-27) ({}).

\bibitem{scannerbit}
G.~D. Martinez, J.~McKay, B.~Farmer, P.~Scott, E.~Roebber, A.~Putze, J.~Conrad,
  {Comparison of statistical sampling methods with ScannerBit, the GAMBIT
  scanning module}, The European Physical Journal C 77~(11) (2017).

\bibitem{gambit}
{The GAMBIT Collaboration}, Gambit: the global and modular
  beyond-the-standard-model inference tool, The European Physical Journal C
  77~(11) (2017) 784.

\bibitem{gambit:page}
{GAMBIT}, \url{https://gambit.hepforge.org}, (2019-06-13) ({}).

\bibitem{chaospy:page}
{chaospy}, \url{https://chaospy.readthedocs.io/en/master/tutorial.html},
  (2020-02-28) ({}).

\bibitem{uncertainpy}
S.~Tenn{\o}e, G.~Halnes, G.~T. Einevoll, {Uncertainpy: A Python Toolbox for
  Uncertainty Quantification and Sensitivity Analysis in Computational
  Neuroscience}, Frontiers in Neuroinformatics 12 (2018).

\bibitem{uncertainpy:page}
{Uncertainpy}, \url{https://uncertainpy.readthedocs.io}, (2019-06-13) ({}).

\bibitem{ghanem2017}
R.~Ghanem, D.~Higdon, H.~Owhadi, Handbook of uncertainty quantification,
  Vol.~6, Springer, 2017.

\bibitem{uqpy:page}
{UQpy}, \url{https://github.com/SURGroup/UQpy}, (2019-06-13) ({}).

\bibitem{Hadjidoukas2012}
P.~E. Hadjidoukas, E.~Lappas, V.~V. Dimakopoulos, {A runtime library for
  platform-independent task parallelism}, Proceedings of the 20th Euromicro
  International Conference on Parallel, Distributed and Network-Based
  Processing (2012) 229--236.

\bibitem{adams2018dakota}
B.~M. Adams, J.~A. Stephens, {Dakota Optimization and UQ: Explore and Predict
  with Confidence.}, Tech. rep., Sandia National Laboratories (2018).

\bibitem{Hansen2016}
N.~Hansen, {The CMA Evolution Strategy: A Tutorial} (2016).
\newblock \href {http://arxiv.org/abs/1604.00772} {\path{arXiv:1604.00772}}.

\bibitem{Hart2003}
W.~E. Hart, {An introduction to the COLIN optimization interface.}, Tech. rep.,
  Sandia National Laboratories (2003).

\bibitem{nair2015massively}
A.~Nair, P.~Srinivasan, S.~Blackwell, C.~Alcicek, R.~Fearon, A.~De~Maria,
  V.~Panneershelvam, M.~Suleyman, C.~Beattie, S.~Petersen, et~al., Massively
  parallel methods for deep reinforcement learning (2015).
\newblock \href {http://arxiv.org/abs/1507.04296} {\path{arXiv:1507.04296}}.

\bibitem{espeholt2018}
L.~Espeholt, H.~Soyer, R.~Munos, K.~Simonyan, V.~Mnih, T.~Ward, Y.~Doron,
  V.~Firoiu, T.~Harley, I.~Dunning, S.~Legg, K.~Kavukcuoglu, Impala: Scalable
  distributed deep-rl with importance weighted actor-learner architectures
  (2018).
\newblock \href {http://arxiv.org/abs/1802.01561} {\path{arXiv:1802.01561}}.

\bibitem{Ching2007}
J.~Ching, Y.~Chen, {Transitional Markov Chain Monte Carlo Method for Bayesian
  Model Updating, Model Class Selection, and Model Averaging}, Journal of
  Engineering Mechanics 133~(7) (2007) 816--832.

\bibitem{MPI:forum}
{MPI Forum}, \url{https://www.mpi-forum.org/} ({}).

\end{thebibliography}

\appendix
\appendixpage

\section{Interface Design} \label{appendix:interface}

\begin{figure}[htb]
\centering
\begin{lstlisting}[style=pythonCode,escapechar=|]
import korali   |\label{code:import:a}|

# Importing the computational model and the data
from myLibrary import F |\label{code:import:b}| 

X = getReferenceInput() |\label{code:get:input}|
Y = getReferenceData()  |\label{code:get:data}|

# Creating new experiment                                                   
e = korali.Experiment() |\label{code:korali:command:experiment}|                           

# Setting up the Bayesian Inference Problem                  
e["Problem"]["Type"] = "Bayesian Inference"  |\label{code:problem:type}|              
e["Problem"]["Likelihood Model"] = "Normal"|\label{code:likelihood:model}|
e["Problem"]["Computational Model"] = lambda s:F(s,X)|\label{code:computational:model}|
e["Problem"]["Reference Data"] = Y  |\label{code:reference:data}|

# Configuring the problem`s variables and their priors
e["Variables"][0]["Name"] = "P1" |\label{code:p1}| 
e["Variables"][1]["Name"] = "P2" |\label{code:p2}|  
e["Variables"][2]["Name"] = "Sigma" |\label{code:sigma}|
e["Variables"][0]["Prior Distribution"] = "D1" |\label{code:p1:prior}|
e["Variables"][1]["Prior Distribution"] = "D1" |\label{code:p2:prior}|
e["Variables"][2]["Prior Distribution"] = "D2" |\label{code:sigma:prior}|

# Configuring the prior distributions
e["Distributions"][0]["Name"] = "D1"  |\label{code:distributions:start}|
e["Distributions"][0]["Type"] = "Univariate/Normal"
e["Distributions"][0]["Mean"]  = 0.0 
e["Distributions"][0]["Sigma"] = +2.0 

e["Distributions"][1]["Name"] = "D2"  
e["Distributions"][1]["Type"] = "Univariate/Uniform"
e["Distributions"][1]["Minimum"] =  0.0 
e["Distributions"][1]["Maximum"] = +5.0  |\label{code:distributions:stop}| 

# Configuring Solver (TMCMC)                                 
e["Solver"]["Type"] = "TMCMC" |\label{code:tmcmc}|
e["Solver"]["Population Size"] = 5000  |\label{code:tmcmc:a}|
e["Solver"]["Covariance Scaling Factor"] = 0.04 |\label{code:solver:covariance}| |\label{code:tmcmc:b}|

# Starting Korali's Engine and running experiment                       
k = korali.Engine() |\label{code:korali:command:engine}|

k.run(e) |\label{code:korali:command:run}|
\end{lstlisting}
\caption{Example of a Python-based Korali Application.}
\label{fig:example:bayesian:inference}
\end{figure}

Korali employs a \textit{descriptive} interface, in which experiments are statically defined as a set of parameters. This interface is mostly language-independent and requires only trivial knowledge of the underlying programming language (\textit{e.g.,} Python or C++). \Cref{fig:example:bayesian:inference} shows an example of a Python-based\footnote{Although we use Python in the examples, Korali provides a similar C++-based interface that allows linking its engine against C++ and Fortran computational models.} application that solves the problem of calibrating the parameters of a computational model on experimental data. We configured the example to describe a Bayesian inference problem where the uncertainty in the parameters is quantified by sampling the posterior distribution of the parameters conditioned on the data. To better explain the software interface, we define the statistical problem first, and then show its correspondence with the code in \cref{fig:example:bayesian:inference}.

The vectors $\vX$ and $\vY$ correspond to the variables \texttt{X} and  \texttt{Y} in the code of \cref{fig:example:bayesian:inference} that are initialized in \cref{code:get:data,code:get:input} through user defined functions.
Korali works by defining and running an \textit{experiment}. (\cref{code:korali:command:experiment})
An experiment consists of the description of a statistical problem to be solved, a description of the involved variables and their distributions, and the configuration of the desired solver.
In this application, all lines of code between \cref{code:problem:type} and \cref{code:solver:covariance} that are required for the description of the problem, are made entirely via dictionary-tree accesses.
The rest of the code consists of importing libraries (\cref{code:import:a,code:import:b}), initializing the experiment (\cref{code:korali:command:experiment}), initializing the engine (\cref{code:korali:command:engine}), and running Korali (\cref{code:korali:command:run}).
The type of the problem is defined in \cref{code:problem:type}, and the likelihood function is defined in \cref{code:likelihood:model}. 
The observations \texttt{Y} are passed to Korali in \cref{code:reference:data} and the computational model in \cref{code:computational:model}. 
In top of \cref{fig:example:models} an example of a computational models is given,
where $f(\X_i; \PM)=\vartheta_1 \X_i + \vartheta_2$.
Next, the variable vector $\PM$ is defined by the experiment's \textit{variables}. Each variable is defined by a unique name and represents one entry to the variable vector.
The example code contains three variables, \texttt{P1}, \texttt{P2} and \texttt{Sigma}, in \cref{code:p1,code:p2,code:sigma}. 
The variables are passed in the user-defined model \texttt{F} and used to compute the likelihood function, given by the keywords \texttt{"Reference Evaluation"} and \texttt{"Standard Deviation"}, respectively. 
To complete the description of the problem, the variables require the definition of a prior distribution $p(\PM)$. Here, we specify that the prior distribution of the variables corresponding to the parameters \texttt{P1} and \texttt{P2} is a normal distribution (\cref{code:p1:prior,code:p2:prior}), and of the variable corresponding to the variable \texttt{Sigma} a uniform distribution (\cref{code:sigma:prior}). Finally, we set the solver to the \texttt{TMCMC} sampler \cite{Ching2007}.

\subsection{Computational Model Support}

The user specifies the computational model by passing a function as part of the problem configuration. Such a function should expect the sample's information as argument. In the example in \cref{fig:example:bayesian:inference}, the computational model is passed as a lambda function that calls the computational model \texttt{F}, imported from the \texttt{myLibrary} module. 

\begin{figure}[htbp]
\centering
\begin{lstlisting}[style=pythonCode]
def F(sample, X): 
    p1 = sample["Variables"]["P1"]
    p2 = sample["Variables"]["P2"]
    s  = sample["Variables"]["Sigma"]
    
    s["Reference Evaluations"] = []
    s["Standard Deviation"]    = []
    for x in X:
        s["Reference Evaluations"] += [a*x + b]
        s["Standard Deviation"]    += [sig]
\end{lstlisting}

\begin{lstlisting}[style=pythonCode]
def myOptimizableModel(sample): 
    x = sample["Variables"]["X"]
    sample["F(x)"] = -x * x
\end{lstlisting}

\begin{lstlisting}[style=pythonCode]
def myExternalModel(sample): 
    x = sample["Variables"]["X"]
    args   = [ './myApp', '-x', str(x) ]
    result = subprocess.check_output(args)
    sample["F(x)"] = float(result)
\end{lstlisting}

\caption{Examples of computational models. (Top): A model that has two parameters \texttt{P1} and \texttt{P2} and produces as result a vector of evaluations, one for each value of the input vector \texttt{X}. (Middle): A model that requires a single variable \texttt{X} and produces a single function evaluation $f(x) = -x^2$, to be maximized using a derivative-free method. (Bottom): A model that executes an external application and returns its output as result.}
\label{fig:example:models}
\end{figure}

Functions passed as computational models do not return a value. Instead, they save their results into the sample container. The expected results from the execution of the computational model depend on the selected problem type. 
\Cref{fig:example:models} (Top) shows the function \texttt{F}, as specified in the example in \cref{fig:example:bayesian:inference}. 
A Bayesian inference problem, where the likelihood is computed from reference data, requires that the model saves an evaluation of each of the reference data points into a \texttt{"Results"} vector. Other problem types, such as \textit{derivative-free optimization}, require the model to store only a single numerical value corresponding to the function evaluation (\texttt{"F(x)"}) for the given parameter(s), as shown in \cref{fig:example:models} (Middle).

The interface accommodates legacy codes through a fork/join-based \texttt{Concurrent} execution mode that allows instancing pre-compiled applications via shell commands and returns the results either through file or pipe I/O operations. The \texttt{Concurrent} mode can also be used to launch and gather results from large-scale distributed, \textit{e.g.,} MPI \cite{MPI:forum} applications. An example of such a model is given in \cref{fig:example:models} (Bottom). 

\section{Experimental Setup} \label{appendix:setup}

We provide here the source code, configuration, and dependencies used to setup the experiments. We performed the experiments on the GPU partition of \pizd \cite{PizDaint2019}, a Cray XC50 system running on SUSE Linux Enterprise Server 15 SP1 (kernel v4.12). For compilation, we used a GNU-based programming environment, with gcc/8.3.0 as C/C++ compiler. We used cray-mpich/7.7.15 for MPI support, and cray-python/3.8.2.1 support Python3. The system uses SLURM 20.02 as its job scheduler. We employed open-source software for the experiments. The software versions used are as follows:
\begin{itemize}
    \item Korali v2.0.0 (\url{github.com/cselab/korali})
    \item $\Pi$4U v1.0 (\url{github.com/cselab/pi4u})
    \item Dakota v6.12 (\url{dakota.sandia.gov})
    \item LAMMPS v20.08 (\url{lammps.sandia.gov})
    \item Mirheo v1.30 (\url{github.com/cselab/Mirheo})
    \item Aphros (\url{github.com/cselab/aphros}).
\end{itemize}

The setup to reproduce the results from \cref{subsection:case1} can be found inside the \path{examples/study.cases/bubblePipe/} folder of the Korali source code; for \cref{subsection:case2}, they can be found inside the \path{examples/study.cases/RBCStretc/} folder, and; for \cref{subsection:case3}, they can be found inside the \path{examples/study .cases/LAMMPS/} folder. The \path{jobs/} folder provides the respective SLURM job scripts used for distributed runs. 

The setup required to run the benchmarks from \cref{section:related} can be found in the Github repository: \url{github.com/cselab/korali_benchmark}. The folders \path{korali/}, \path{pi4u/}, and \path{dakota/} contain the configuration used for each of the frameworks tested. The code files in \path{korali/} and \path{pi4u/} need to be compiled with \path{make} prior to running the scripts. Inside the \path{common/} folder is the common wait script for the three frameworks to use during sample evaluations. The scripts within the \path{jobs/} folder contain variables that allow setting the experiment's workload imbalance.

\end{document}